


\magnification=\magstep1
\baselineskip=12pt
\overfullrule=0pt
\hsize = 5.5 in
\vsize = 7.8 in
\centerline{\bf Symplectic and Poisson Geometry on Loop Spaces}
\centerline{\bf of Manifolds and Nonlinear Equations}
\vskip .4 in
\centerline{\bf Oleg Mokhov}
\vskip .2 in
\centerline{\it Department of Geometry and Topology}
\centerline{\it Steklov Mathematical Institute}
\centerline{\it ul. Vavilova, 42}
\centerline{\it Moscow, GSP-1, 117966, Russia}
\centerline{\it e-mail: mokhov@class.mian.su;  mokhov@top.mian.su}
\vskip .6 in
We consider some differential geometric classes of local and nonlocal
Poisson and symplectic structures on loop spaces of smooth manifolds
which give natural Hamiltonian or multihamiltonian representations
for some important nonlinear equations of mathematical physics and
field theory such as nonlinear sigma models with torsion, degenerate
Lagrangian systems of field theory, systems of hydrodynamic type,
N-component systems of Heisenberg magnet type, Monge-Amp\`ere equations,
the Krichever-Novikov equation
and others. In particular, complete classification of all
nondegenerate Poisson bivectors $\omega ^{ij}(x,u,u_x,u_{xx},...)$
depending on derivatives of the field variables $u^i (x)$
and the independent space variable $x$ is obtained ($u^i,\  i=1,...,N,$
are local coordinates on smooth manifold $M$).
In other words, all Poisson brackets of the following form
$$\{ u^i (x), u^j (y) \} = \omega ^{ij} (x, u, u_x, u_{xx}, ...)
\delta (x-y), $$
$\det ( \omega ^{ij} ) \neq 0$, are explicitly described.
In addition, we shall prove integrability of some class of
nonhomogeneous systems of hydrodynamic type and give a description
of nonlinear partial differential equations of
associativity in $2D$ topological
field theories (for some special type solutions of
the Witten-Dijkgraaf-E.Verlinde-H.Verlinde (WDVV) system)
 as integrable nondiagonalizable weakly nonlinear homogeneous system
of hydrodynamic type.

\vfill\eject

\noindent{\bf 1  Generalized Poisson bivectors on loop spaces
of manifolds}

\vskip .2 in
Let be given an $N$-dimensional smooth manifold $M$ with local
coordinates $\{ u^1,...,u^N \}$. Classic Poisson bivector
$\omega ^{ij} (u)$ on $M$ is by definition a skew-symmetric
(2,0)-tensor on $M$ ($\omega ^{ij} (u) = - \omega ^{ji} (u)$)
which satisfies the well-known relation (the Jacobi identity):
$${\partial \omega ^{ij} \over \partial u^s} \omega ^{sk} +
{\partial \omega ^{jk} \over \partial u^s} \omega ^{si} +
{\partial \omega ^{ki} \over \partial u^s} \omega ^{sj} = 0
   \eqno{(1.1)}$$
It gives a Poisson bracket on the space of smooth functions on $M$:
$$\{ f(u), g(u) \} = \omega ^{ij} (u) {\partial f \over
\partial u^i} {\partial g \over \partial u^j}  \eqno{(1.2)}$$
(or $\{ u^i, u^j \} = \omega ^{ij} (u)$) which is
skew-symmetric $$\{f(u), g(u)\} = - \{g(u), f(u)\}$$
 and satisfies the Jacobi identity
$$\{f(u),\{g(u), h(u)\}\} + \{g(u),\{h(u), f(u)\}\} +
 \{h(u), \{f(u), g(u)\}\} = 0$$

First of all, in this paper  we would like to study some natural
infinite-dimensional generalizations of the Poisson bivectors
and the corresponding Poisson brackets.

Let us consider the
loop space of the manifold $M$, i.e. the space  $L(M)$
of all smooth parametrized mappings $\gamma : S^1 \rightarrow
M,\  \gamma (x) = \{ u^i (x),\  x \in S^1 \}$.
The classic Poisson bivector $\omega ^{ij} (u)$ gives so-called
ultra-local Poisson bracket on the loop space $L(M)$:
$$\{ u^i (x), u^j (y) \} = \omega ^{ij} (u(x)) \delta (x-y)  \eqno{(1.3)}$$
or, in other words,
we have the Poisson bracket on the space of functionals on the loop space
$L(M)$
$$\{  F, G \} = \int_{S^1} {\delta F \over \delta u^i (x)}
\omega ^{ij} (u(x)) {\delta G \over \delta u ^j (x)} dx  \eqno{(1.4)}$$
where $F$ and $G$ are arbitrary functionals on $L(M)$.
In infinite-dimensional case of $L(M)$ we can consider a natural
generalization of the Poisson bracket (1.4).
In fact, generally speaking, in this case a Poisson bivector $\omega ^{ij}$
can depend also on derivatives of the fields $u^k (x)$:
$\omega ^{ij} (u, u_x, u_{xx},...)$.

\noindent{\it Problem 1}. Describe all Poisson brackets on $L(M)$ of
the following form:
$$\{ F, G \} = \int_{S^1} {\delta F \over \delta u^i (x)}
\omega ^{ij} (u, u_x, u_{xx},...) {\delta G \over \delta u^j (x)} dx
\eqno{(1.5)}$$
where $\det ( \omega ^{ij}) \neq 0$.

\noindent{\it Remark 1.1}. Note that in this paper we always consider
only tensors depending on a finite number of derivatives of the fields
$u^k(x)$.

\noindent{\it Definition 1.1}. We shall call a bivector $\omega ^{ij}
(u, u_x, u_{xx},...)$ by a generalized Poisson bivector on
the loop space $L(M)$ if the tensor
$\omega ^{ij} (u, u_x, u_{xx},...)$ gives a Poisson bracket (1.5)
on the space of functionals on $L(M)$.

\noindent{\bf Theorem 1.1}. Nondegenerate tensor
 $\omega ^{ij} (u, u_x, u_{xx},...)$,
$\det (\omega ^{ij}) \neq 0$, on the loop space $L(M)$ is a generalized
Poisson bivector on $L(M)$ if and only if there are a closed 2-form
$\Omega _{ij} (u)$ on $M$ ($d \Omega = 0$) and a closed 3-form
$T_{ijk} (u)$ on $M$ ($dT = 0$) such that $\det (T_{ijk} (u) u^k_x +
\Omega _{ij} (u)) \neq 0$ and
$$\omega ^{ij} (T_{jks} (u) u^s_x + \Omega _{jk} (u)) = \delta ^i_k
\eqno{(1.6)} $$

\noindent{\it Remark 1.2}. Thus, it follows from (1.6) that
nondegenerate generalized Poisson bivectors $\omega ^{ij}$
on $L(M)$ can depend only on the first derivatives of the field
variables $u^i (x)$.

\noindent{\it Remark 1.3}. If the manifold $M$ is a symplectic one
and the 2-form $\Omega _{ij} (u)$ is a symplectic form on $M$
(i.e. $\det (\Omega _{ij}) \neq 0$) then for any closed
3-form $T_{ijk} (u)$ on $M$ we have always that $\det (T_{ijk} (u) u^k_x
+ \Omega _{ij} (u)) \neq 0$ and the formula (1.6) gives explicitly
a generalized Poisson bivector $\omega ^{ij}$.
We note also here that for any closed 3-form $T_{ijk}$ on $M \ $ $\det (T_{ijk}
(u) u^k_x) \equiv 0$.

 Theorem 1.1 describes translation
invariant Poisson brackets but we can consider also Poisson bivectors
$\omega ^{ij} (x, u(x), u_x, u_{xx},...)$ depending explicitly on
the independent space variable $x$.

\noindent{\bf Theorem 1.2}. Tensor $\omega ^{ij} (x, u, u_x, u_{xx},...)$,
$\det (\omega ^{ij}) \neq 0$, gives a Poisson bracket
$$\{ F, G \} = \int_{S^1} {\delta F \over \delta u^i (x)}
\omega ^{ij} (x, u, u_x, u_{xx},...) {\delta G \over \delta u^j (x)} dx$$
on the space of functionals on $L(M)$ if and only if there are an
one-parameter family of 2-forms $\Omega _{ij} (z, u)$
on the manifold $M$ and a closed 3-form $R_{ijk} (u)$ on $M$ such that
$$\det [( \int^x_0 (d \Omega)_{ijk}(z,u(x)) dz + R_{ijk} (u)) u^k_x +
\Omega_{ij} (x, u(x))] \neq 0$$ and
$$\omega ^{is} [( \int^x_0 (d \Omega)_{sjk} (z, u(x)) dz +
R_{sjk} (u)) u^k_x + \Omega _{sj} (x, u(x))] = \delta ^i_j, \eqno{(1.7)}$$
where
$$(d \Omega)_{ijk} (z, u(x)) = {\partial \Omega _{ij} (z, u(x)) \over
\partial u^k}  + {\partial \Omega _{ki} (z, u(x)) \over \partial u^j}
+ {\partial \Omega _{jk} (z, u(x)) \over \partial u^i}$$
(note that there are no conditions of closure for the 2-forms
$\Omega _{ij}$ here as it takes place in the Theorem 1.1).

\noindent{\it Remark 1.4}. It is an interesting unsolved problem
to classify all degenerate generalized Poisson bivectors
$\omega ^{ij} (x,u,u_x,u_{xx},...)$ with respect to dependence on
derivatives of the fields $u^i(x)$.

In order that to prove the Theorems 1.1 and 1.2 we shall consider
differential geometric objects which are inverse to nondegenerate
Poisson bivectors $\omega ^{ij}$. As it is well known,
in finite-dimensional case we shall get
 symplectic forms $\omega _{ij} (u)$
on the manifold $M$: $\omega _{ij} (u) \omega ^{jk} (u) = \delta ^k_i$,
$(d \omega )_{ijk} = 0$. In the next section we shall
consider corresponding generalizations of symplectic structures
in connection with infinite-dimensional case
of the loop space $L(M)$.
\vskip .3 in

\noindent{\bf 2  Symplectic structures on loop spaces of manifolds}

\vskip .2 in
The tangent space of the loop space $L(M)$ in its point (a loop) $\gamma$
consists of the all smooth vector fields $\xi ^i, 1 \leq i \leq N$,
defined along the loop $\gamma$. We shall denote it here by
$T_{\gamma} L(M)$ (we note that $\xi ^i (\gamma (x)) \in T_{\gamma (x)} M$,
$\ \forall x \in S^1$, where $T_{\gamma (x)} M$ is the tangent space of the
manifold $M$ in the point $\gamma (x)$).

Consider a skew-symmetric bilinear form $\omega (\xi, \eta)$ on the loop
space $L(M)$:
$$ \omega (\xi, \eta) = \int _{S^1} \xi ^i A_{ij} \eta ^j dx,
\eqno{(2.1)} $$
$\omega (\xi, \eta) = - \omega (\eta, \xi)$,
$\ \xi, \eta \in T_{\gamma}L(M),$
$\ \xi = (\xi^1,...,\xi^N),\ \eta = (\eta^1,...,\eta^N),\ $
 $A = (A_{ij})$ is a skew-symmetric operator defined by the loop
$\gamma$
$$A[\gamma] : T_{\gamma} L(M) \rightarrow T^*_{\gamma} L(M)$$

Introduce a differential of 2-form $\omega(\xi, \eta)$
(it is an natural infinite-dimensional
generalization of the usual differential
in the Lie algebra of vector fields on a manifold):
$$ (d \omega ) (\xi, \eta, \zeta) = \sum _{(\xi, \eta, \zeta)}
\biggl \{ \int _{S^1} \xi ^i {\delta \omega (\eta, \zeta) \over
\delta u^i} dx + \omega (\xi, [\eta, \zeta]) \biggr \},  \eqno{(2.2)}$$
where
$\ \xi,\eta,\zeta \in T_{\gamma}L(M),$
$$[\eta, \zeta]^j = \eta ^i_{(k)} {\partial \zeta ^j \over
\partial u^i_{(k)}} - \zeta ^i_{(k)} {\partial \eta ^j \over
\partial u^i_{(k)}}$$
is a commutator of the vector fields $\eta$ and $\zeta$,
$f_{(k)} = {d^k f / d x^k}$, and the sign $\sum _{(\xi, \eta,
\zeta)}$ means that the sum is taken with respect to all cyclic
permutations of the elements $(\xi, \eta, \zeta)$.

If $d \omega = 0$, i.e. the 2-form (2.1) is closed, then
the operator $A_{ij}$ is called by a
symplectic operator ($\omega $ is called by presymplectic
form or presymplectic structure).

\noindent{\it Remark 2.1}. In infinite-dimensional case
presymplectic structure is often said to be a symplectic one
even if the corresponding
2-form $\omega$ is degenerate on $L(M)$. We shall keep this terminology here
(see [30] for refs. on symplectic operators and general theory of
local Hamiltonian and symplectic structures and their connections with
integrability of nonlinear partial differential equations).

\noindent{\bf Lemma 2.1}. Nondegenerate tensor
$\omega^{ij}(x,u,u_x,u_{xx},...)$, $\ \det \omega ^{ij} \neq 0$,
is a generalized Poisson bivector on $L(M)$ if and only if
the tensor $\omega _{ij} (x,u,u_x,u_{xx},...)$, where
$\omega ^{is} \omega _{sj} = \delta ^i_j$, is a symplectic operator.

It was shown by author in
[13] that it is sufficient to verify the Jacobi identity only for
all linear functionals of the form $F = \int _{S^1} f_i(x) u^i(x)dx$,
where $f_i(x),\   i = 1,...,N$, are arbitrary functions.
Consider vector fields $\xi$ of the form $\xi ^i = \omega ^{ij} f_j(x)$.
It is easy to check that the condition
$$(d \omega)(\xi,\eta,\zeta) = 0$$
for a symplectic structure $\omega _{ij} (x,u,u_x,u_{xx},...)$
is equivalent to the Jacobi identity on linear functionals for the
bracket
$$\{u^i(x),u^j(y) \} = \omega ^{ij} (x,u,u_x,u_{xx},...) \delta (x-y).$$

Consider now symplectic operators $\omega _{ij} (u, u_x, u_{xx},...)$.

\noindent{\bf Lemma 2.2}. An operator $\omega _{ij} (u, u_x, u_{xx},...)$
is a symplectic operator if and only if

1) $\omega _{ij} = - \omega _{ji} $ (skew-symmetry)

2) for any vector-functions $\eta (x) = (\eta ^1 (x),...,\eta ^N (x)),
\zeta (x) = (\zeta ^1 (x),...,\zeta ^N (x))$ we have the following
identity:
$$ (-1)^k \biggl ( {d \over dx} \biggr ) ^k  \biggl ( {\partial
\omega _{ij} \over \partial u^s_{(k)} } \eta ^i (x) \zeta ^j (x)
\biggr ) + {\partial \omega _{si} \over \partial u^j _{(k)}}
\eta ^i (x) \zeta ^j_{(k)} (x) - { \partial \omega _{sj} \over
\partial u^i_{(k)}} \eta ^i_{(k)} (x) \zeta ^j (x) \equiv 0
\eqno{(2.3)}$$
(the condition of closure for $\omega _{ij}$).

Consider now
$ R = \max \biggl \{ m:$ there exist $i,j,s$ such that
${\partial \omega _{ij}} / {\partial u^s_{(m)}} \neq 0 \biggr \}$.

If $R \geq 2$ then the coefficient at the term $\eta ^i_x (x)
\zeta ^j _{(R-1)} (x)$ in (2.3) is equal $$(-1)^R R {\partial
\omega _{ij} \over \partial u^s_{(R)}},$$
i.e. $R \leq 1$ and the form $\omega_{ij}$ depends only on $u^k (x)$ and the
first
derivatives $u^k_x (x)$.

Now it is easy to show from the identity (2.3) that $\omega _{ij}$
must be quasilinear with respect to the first derivatives $u^k_x (x)$,
i.e.
$$ \omega _{ij} (u, u_x) = T_{ijk} (u) u^k_x + \Omega _{ij} (u)
\eqno{(2.4)}$$
Note that all symplectic operators of this type were described
by author in [8] where differential geometric homogeneous
symplectic operators of the first order $$A_{ij} = g_{ij}
{d \over dx} + b_{ijk} (u) u^k_x$$ were studied in detail
(after degeneration of $g_{ij}$ we get complete description
of quasilinear symplectic structures $\omega _{ij} (u, u_x)$).

It turns out that symplectic operators $\omega _{ij}$ of the form (2.4)
give Hamiltonian (or symplectic)
structures of some special Lagrangian systems in
classical field theory (some degeneration of nonlinear sigma models).

\noindent{\bf Theorem 2.1}. Lagrangian systems $\delta S / \delta u =0$
that are given by the action
$$S = \int (a_{ij} (u) u^i_x u^j_t + b_i (u) u^i_t + U(x,t,u, u_x, u_{xx},
...)) dx dt \eqno{(2.5)}$$
with $b_i (u), U (x,t,u, u_x, u_{xx},...)$ being arbitrary functions and
skew-symmetric functions $a_{ij} (u) $ ( $a_{ij} = - a_{ji} $),
are Hamiltonian systems with respect to the corresponding symplectic
structures of the form (2.4).

Conversely, any Hamiltonian system of the form
$$(T_{ijk} (u) u^k_x + \Omega _{ij} (u)) u^j_t =
{\delta U \over \delta u^i}$$
is Lagrangian system with respect to the action of the type (2.5).

In the next section we shall consider symplectic geometry corresponding
to the general nonlinear sigma models with torsion.

\vskip .3 in
\noindent{\bf 3  Nonlinear sigma models with torsion and
symplectic geometry on loop spaces of manifolds}

\vskip .2 in
Consider two-dimensional systems generated by the general nonlinear
sigma model actions of the form
$$S = \int ( {1 \over 2} r_{ij} (u) u ^i_x u^j_t + U(u) ) dxdt
\eqno{(3.1)}$$
where $r_{ij} (u)$ is an arbitrary (0,2)-tensor and $U(u)$ is an
arbitrary function on $M$.
As it was shown by author in [8] the corresponding Lagrangian system
$${\delta S \over \delta u} = 0 $$ has always symplectic representation
$$A_{ij} u^j_t = {\delta U \over \delta u}  \eqno{(3.2)}$$
where
$$A_{ij} = {1 \over 2 } ( r_{ij} + r_{ji}) {d \over dx} + {1 \over 2}
\biggl ( {\partial r_{ki} \over \partial u^j} + {\partial r_{ij}
\over \partial u^k} - {\partial r_{kj} \over \partial u^i} \biggr )
u^k_x  \eqno{(3.3)}$$
is a symplectic operator.

If $\det (r_{ij} + r_{ji}) \neq 0$
(in this case we have Riemannian or pseudoRiemannian manifold $M$
with the metric $g_{ij} = {1 \over 2} (r_{ij} + r_{ji})$)
then the symplectic operator (3.3) gives a symplectic form
$\omega (\xi, \eta) $ on $L(M)$ having the shape
$$\omega (\xi,\eta) = \int _{\gamma} < \xi, \nabla _{\dot{\gamma}}
\eta >   \eqno{(3.4)}$$
where $< \xi, \eta> = {1 \over 2} ( r_{ij} + r_{ji}) \xi ^i \eta ^j$
and $\nabla _{\dot{\gamma}}$ is a covariant derivation along
$\gamma \ $ generated by
some differential geometric connection $\Gamma ^i_{jk} (u)$ on $M$.
Let us describe all symplectic forms of the shape (3.4) on $L(M)$.

\noindent{\bf Theorem 3.1  [8]}. Let be given a Riemannian or
pseudoRiemannian manifold $(M, g_{ij})$. A differential geometric
connection $\Gamma ^i_{jk} (u)$ on $M$ defines a symplectic form
(3.4), where $< \xi, \eta > = g_{ij} \xi ^i \eta ^j$, if and only if

1) the connection $\Gamma ^i_{jk} (u)$ is compatible with the metric
$g_{ij}$, i.e.
$$\nabla _k g_{ij} \equiv {\partial g_{ij} \over \partial u^k}
- g_{is} \Gamma ^s_{jk} - g_{js} \Gamma ^s_{ik} = 0$$

2) the torsion $$T_{ijk} = g_{is} T^s_{jk},\   T^i_{jk} \equiv
\Gamma ^i_{jk} - \Gamma ^i_{kj},$$
is a closed 3-form on $M$.

Any closed 3-form on the (pseudo)Riemannian manifold $(M, g_{ij})$
gives the only compatible with the metric $g_{ij} (u)$ differential
geometric connection $\Gamma ^i_{jk} (u)$ with torsion tensor
determined explicitly by this closed 3-form and correspondingly
any metric $g_{ij} (u)$ and any closed 3-form on $M$ generate
a symplectic form (3.4) on $L(M)$.

The corresponding symplectic operator has the form
$$M_{ij} = g_{ij} (u) {d \over dx} + g_{is} (u) \Gamma ^s_{jk} (u)
u^k_x   \eqno{(3.5)}$$
where locally we have
$$\Gamma ^i_{jk} (u) ={1 \over 2} g^{is} (u) \biggl (
{\partial g_{sk} \over \partial u^j} + {\partial g_{js} \over
\partial u^k} - {\partial g_{jk} \over \partial u^s}
+ T_{sjk} (u) \biggr ),   \eqno{(3.6)}$$

$$T_{ijk} (u) = {1 \over 2} \biggl ( {\partial f_{ij} \over
\partial u^k} + {\partial f_{jk} \over \partial u^i} +
{\partial f_{ki} \over \partial u^j} \biggr ),  \eqno{(3.7)}$$

$$f_{ij} (u) = - f_{ji} (u)  \eqno{(3.8)}$$

\noindent{\it Corollary}. On any Riemannian or pseudoRiemannian
manifold $(M, g_{ij})$ closed 2-forms of the shape (3.4) on $L(M)$
are in one-to-one correspondence with closed 3-forms on $M$.

An elementary conclusion is that on any two-dimensional Riemannian
(or pseudoRiemannian) manifold $(M, g_{ij})$ there is a single
symplectic 2-form (3.4) on the loop space $L(M)$
and correspondingly a single symplectic operator (3.5)
which are generated by the
Levi-Civita connection.  As for $N$-dimensional Riemannian
(or pseudoRiemannian) manifolds, the conclusion is that there
always exist an symplectic 2-form of the shape (3.4)
and a symplectic operator (3.5) which are
generated by the Levi-Civita connection.

It must be noted that the symplectic form (3.4) is always
degenerated (i.e. in fact, it is only a presymplectic form on $L(M)$)
and the null space of $\omega (\xi, \eta)$
is constituted by vector fields parallel along $\gamma$.
In particular, the velocity vector field $\{ u^i_x \}$ belongs
to the null space of $\omega (\xi, \eta)$ if $\gamma$ is a
geodesic loop. Another subject is the interrelation between
the symplectic forms under consideration and classic finite-dimensional
symplectic structures on manifolds of geodesics.
We remind that for $N$-dimensional Riemannian manifolds
$(M, g_{ij})$ whose geodesics are periodic and of equal length,
there exists a $(2N-2)$-dimensional symplectic manifold
$CM$ of geodesics. The symplectic structure on $CM$ is determined
by the curvature form of $S^1$-connection
in the principal bundle $UM$ of unit tangent vectors of
$(M, g_{ij})$, where the $S^1$-connection is generated by the
canonical 1-form on $T^{\ast}M$ (Reeb theorem, see [33--36]).
The tangent space $T_{\gamma} CM$ at the point $\gamma$
is isomorphic to the space of normal
Jacobian fields along the geodesic $\gamma$.

\noindent{\bf Theorem 3.2 [8]}. The restriction of the
symplectic form (3.4) generated on $L(M)$ by the Levi-Civita
connection to the finite-dimensional subspace of normal
Jacobian fields along the geodesics $\gamma$ coincides
with the symplectic Reeb form that is a closed nondegenerate
2-form on $CM$ defining a symplectic structure.

\noindent{\bf Theorem 3.3 [8]}. Formulas (3.5)-(3.8)
give complete description of all symplectic operators of the
following shape
$$A_{ij} = a_{ij} (u) {d \over dx} + b_{ijk} (u) u^k_x,
\eqno{(3.9)}$$
$$\det a_{ij} (u) \neq 0$$
(homogeneous differential operators of the first order with
nondegenerate higher coefficient).

\noindent{\bf Lemma 3.1}. An operator of the shape (3.9)
is symplectic if and only if the following conditions are valid:

$$a_{ij} = a_{ji}     \eqno{(3.10)}$$

$${\partial a_{ij} \over \partial u^k} = b_{ijk} + b_{jik}   \eqno{(3.11)}$$

$${\partial a_{ij} \over \partial u^k} = b_{ikj} + b_{jki}   \eqno{(3.12)}$$

$${\partial b_{jmk} \over \partial u^i} - {\partial b_{jmi} \over
 \partial u^k} + {\partial b_{ijk} \over \partial u^m} -
{\partial b_{imk} \over \partial u^j} =0   \eqno{(3.13)}$$

\noindent{\it Remark 3.1}. Hamiltonian operators of the shape (3.9)
were classified by Dubrovin and Novikov [2] (see also [1,4]).
In this case, the higher coefficient $a^{ij} $ must be a flat
metric on manifold $M$ and $b^{ij}_k (u)$ are generated by this
flat metric. These Hamiltonian operators play the very important role
in the theory of integrability of the diagonalizable systems
of hydrodynamic type
[5,6] (see also [1,4]).

Thus, any nonlinear sigma model (3.1) has a natural Hamiltonian
(symplectic) representation (3.2) given by a symplectic operator (3.3).
If a nonlinear sigma model system has also another Hamiltonian
representation compatible with (3.2) then this nonlinear sigma model
system will be
integrable by the Lenart-Magri scheme ([37]). Thus, the problem
of finding of Hamiltonian operators compatible with the symplectic
operator (3.3) is very important and was formulated by author in 1989
(see [8,11,38--40]). For some known two-component integrable
nonlinear sigma models without torsion ($r_{ij}$ is a symmetric tensor
in (3.1)) such second Hamiltonian representations compatible with (3.2), (3.3)
were found in [41].

\noindent{\it Example 3.1} [41].
$$S = \int \bigl ( {1 \over 2} {{u^1_x u^2_t + u^2_x u^1_t} \over
{u^1 u^2 + c}} + k u^1 u^2 \bigr )dxdt \eqno{(3.14)}$$
where $c$ and $k$ are arbitrary constants.

In this case
$$(r_{ij}) ={1 \over {u^1 u^2 + c}} { \pmatrix {0 & 1  \cr
                         1  & 0 \cr}},$$
$$U(u) = k u^1 u^2$$.

First symplectic operator (3.3) has the form

$$A={1 \over {u^1 u^2 + c}} {\pmatrix {0 & D \cr
                          D & 0 \cr }} -
{1 \over {(u^1 u^2 + c)^2}} {\pmatrix {0 & u^1 u^2_x \cr
                             u^2 u^1_x & 0 \cr }} \eqno{(3.15)}$$
The second Hamiltonian operator $B$ for the action (3.14) is
nonlocal and compatible with (3.15):
$$B = {\pmatrix { 0 & -(u^1 u^2 +c) \cr
                  (u^1 u^2 +c) & 0 \cr}} +$$
$${\pmatrix {u^1_x D^{-1} \circ u^1 & -u^1_x D^{-1} \circ u^2 \cr
           u^2_x D^{-1} \circ u^1 & -u^2_x D^{-1} \circ u^2 \cr}} +
{\pmatrix {u^1 D^{-1} \circ u^1_x & u^1 D^{-1} \circ u^2_x \cr
           -u^2 D^{-1} \circ u^1_x & -u^2 D^{-1} \circ u^2_x \cr}}
\eqno{(3.16)} $$

The recursion operator $L^i_j = B^{is}A_{sj}$ for this integrable
nonlinear sigma model (3.14) has the following form
$$L=
{\pmatrix {-D & 0 \cr
            0 & D \cr}} + {1 \over {u^1 u^2 + c}}
{\pmatrix {u^1 u^2_x & 2 u^1 u^1_x \cr
           -2 u^2 u^2_x & - u^2 u^1_x \cr}} +$$
$${\pmatrix {u^1_x D^{-1} \circ {cu^2_x \over {(u^1 u^2 +c)^2}} &
             u^1_x D^{-1} \circ {-cu^1_x \over {(u^1 u^2 +c)^2}} \cr
             u^2_x D^{-1} \circ {cu^2_x \over {(u^1 u^2 +c)^2}} &
             u^2_x D^{-1} \circ {-cu^1_x \over {(u^1 u^2 +c)^2}} \cr}} +$$
$${\pmatrix {u^1 D^{-1} \circ \bigl ( {u^1 (u^2_x)^2
\over {(u^1 u^2 +c)^2}} -{u^2_{xx} \over {u^1 u^2 +c}} \bigr ) &
           u^1 D^{-1} \circ \bigl ( {u^2 (u^1_x)^2
\over {(u^1 u^2 +c)^2}} -{u^1_{xx} \over {u^1 u^2 +c}} \bigr ) \cr
          -u^2 D^{-1} \circ \bigl ( {u^1 (u^2_x)^2
\over {(u^1 u^2 +c)^2}} -{u^2_{xx} \over {u^1 u^2 +c}} \bigr ) &
          -u^2 D^{-1} \circ \bigl ( {u^2 (u^1_x)^2
\over {(u^1 u^2 +c)^2}} -{u^1_{xx} \over {u^1 u^2 +c}} \bigr ) \cr}}
\eqno{(3.17)}$$

\noindent{\it Example 3.2} [41].
$$S= \int \bigl ( {1 \over 2} {{u^1_x u^2_t + u^2_x u^1_t} \over
{u^1 +u^2} } + k (u^1 +u^2) \bigr )dxdt \eqno{(3.18)}$$
where $k$ is an arbitrary constant.

For this case
$$(r_{ij}) = {1 \over {u^1 + u^2}}
{\pmatrix {0 & 1 \cr
          1 & 0 \cr}},$$
$$U(u) = k(u^1+u^2)$$.

The first symplectic operator (3.3) has the following form in this case:

$$A={1 \over {u^1 +u^2}} {\pmatrix {0 & D \cr
                           D & 0 \cr}} -{1 \over {(u^1 + u^2)^2}}
{\pmatrix {0 & u^2_x \cr
           u^1_x & 0 \cr}}  \eqno{(3.19)}$$

The second Hamiltonian operator $B$ for the action (3.18) is
also nonlocal and compatible with (3.19):
$$B= {\pmatrix {0 & -(u^1+u^2) \cr
              (u^1 +u^2) & 0 \cr}} +$$
$${\pmatrix {u^1_x D^{-1}  & -u^1_x D^{-1} \cr
             u^2_x D^{-1} & -u^2_x D^{-1} \cr}}
+{\pmatrix {D^{-1} \circ u^1_x & D^{-1} \circ u^2_x \cr
            -D^{-1} \circ u^1_x & - D^{-1} \circ u^2_x \cr}}
\eqno{(3.20)}$$

The recursion operator $L^i_j = B^{is} A_{sj} $ has the form

$$L= {\pmatrix {- D & 0 \cr
                 0 & D \cr}} + {1 \over {u^1 + u^2}}
{\pmatrix {u^2_x & 2u^1_x \cr
         - 2 u^2_x & - u^1_x \cr}} +$$
$${\pmatrix {-u^1_x D^{-1} \circ { u^2_x \over {(u^1 +u^2)^2}} &
u^1_x D^{-1} \circ {u^1_x \over {(u^1 +u^2)^2}} \cr
-u^2_x D^{-1} \circ {u^2_x \over {(u^1 + u^2)^2}} &
u^2_x D^{-1} \circ {u^1_x \over {(u^1 + u^2)^2}} \cr}} +$$
$${\pmatrix { D^{-1} \circ \biggl ( {(u^2_x)^2 \over
{(u^1 +u^2)^2}} - {u^2_{xx} \over {u^1 + u^2}} \biggr ) &
D^{-1} \circ \biggl ( {{(u^1_x)^2} \over {(u^1 + u^2)^2}} -
{{u^1_{xx}} \over {u^1 + u^2}} \biggr )
\cr
D^{-1} \circ \biggl ( -{{(u^2_x)^2} \over {(u^1 + u^2)^2}} +
{u^2_{xx} \over {u^1 +u^2}} \biggr ) &
D^{-1} \circ \biggl ( - {{(u^1_x)^2} \over (u^1 + u^2)^2} +
{u^1_{xx} \over {u^1 + u^2}} \biggr ) \cr }} \eqno{(3.21)}$$

\vskip .3 in
\noindent{\bf 4  Homogeneous symplectic forms on loop spaces of
manifolds}

\vskip .2 in
Consider the matrix entries
of the operator $A^{[m]}$ of the shape
$$A^{[m]}_{ij} = a^{[m]}_{ij} (u) {d^m \over dx^m} +
b^{[m]}_{ijk} (u) u^k_x {d^{m-1} \over dx^{m-1}} +
(c^{[m]}_{ijkl}(u) u^k_x u^l_x + d^{[m]}_{ijk} (u) u^k_{xx})
{d^{m-2} \over dx^{m-2}} + ...,   \eqno{(4.1)}$$
with all the terms in (4.1) being degree m homogeneous with
respect to the natural grading
$$\deg (fg) =\deg f + \deg g,$$
$$ \deg f(u(x)) = \deg u(x) =0, $$
$$\deg {d^k u \over dx^k} = \deg {d^k \over dx^k} = k $$

 Symplectic operators of the shape (4.1)
and the corresponding symplectic forms on the loop space $L(M)$
$$\omega (\xi, \eta) = \int _{S^1} \xi ^i (A^{[m]}_{ij} \eta ^j) dx
 \eqno{(4.2)}$$
were introduced and studied
by author in [9,38,39].

It can be proved easily that for symplectic form (4.2) the coefficients of
the operator (4.1) are transformed under changes of coordinates $u^i =
u^i (v)$ on the $N$-dimensional manifold $M$ as differential geometric
objects.
Under the assumption of nondegeneracy of the higher coefficient
$a^{[m]}_{ij}(u)$ the conditions reflecting symplecticity of the
operator (4.1) are in their essence some natural differential-geometric
restrictions imposed on connections defined on manifold
$M$ equipped with the metric $a^{[m]}_{ij} $ (symmetric for
any odd $m$,  $\ m= 2k+1$, or skew-symmetric for any even $m$,
$\ m=2k$)
and the coefficients can be expressed in an invariant form via
the metric, the curvatures and the torsions of the connections.
It allows to obtain invariant geometric description of the
symplectic operators (4.1).

\noindent{\it Definition 4.1}. Symplectic operators of the shape
 (4.1) are called symplectic operators of differential geometric
type.

\noindent{\it Remark 4.1}. Poisson structures of the shape (4.1)
were introduced by Dubrovin and Novikov [3] as a natural generalization
of Poisson structures of hydrodynamic type ($m=1$). The case $m=2$
for Poisson structures of the shape (4.1) was studied in [44--46].

For $m = 0$  symplecticity conditions are
$$a^{[0]}_{ij} (u) = - a^{[0]}_{ji} (u),$$

$$\sum _{(i,j,k)} {\partial a^{[0]}_{ij} \over \partial u^k} = 0,$$
where the sum is taken with respect to all cyclic permutations of $i,j,k$.
It means that
$$\Omega _0 = a^{[0]}_{ij} (u) du^i \wedge du^j$$
is an arbitrary closed 2-form on $M$. If $\det a^{[0]}_{ij} \neq 0$,
we have a symplectic structure $\Omega _0$ on $M$ and a symplectic
structure
$$\omega (\xi, \eta) = \int _{S^1} \Omega _0 (\xi, \eta) dx$$
on the loop space $L(M)$.

The case $m = 1$ corresponds exactly to symplectic forms (3.4).
It is possible also to get a complete description of matrix
symplectic operators (4.1) for $ m = 2$ [9]. In this case we deal
with the operators of the form
$$A^{[2]}_{ij} = a^{[2]}_{ij} (u) {d^2 \over dx^2} +
b^{[2]}_{ijk} (u) u^k_x {d \over dx}
+c^{[2]}_{ijkl} (u) u^k_x u^l_x + d^{[2]}_{ijk} (u) u^k_{xx}  \eqno{(4.3)}$$

\noindent{\bf Lemma 4.1}. Operator (4.3) is symplectic if and only if
the following conditions are valid:
$$a^{[2]}_{ij} = - a^{[2]}_{ji},     \eqno{(4.4)}$$
$${\partial a^{[2]}_{ij} \over \partial u^k } = {1 \over 2}
(b^{[2]}_{ijk} - b^{[2]}_{jik}),    \eqno{(4.5)} $$
$$d^{[2]}_{ijk} = {1 \over 2} ( b^{[2]}_{ijk} + b^{[2]}_{kji}
- b^{[2]}_{kij}),  \eqno{(4.6)}$$
$$c^{[2]}_{ijkl} + c^{[2]}_{ijlk} = {\partial
d^{[2]}_{ijk} \over \partial u^l} + {\partial d^{[2]}_{ijl}
\over \partial u^k} + {\partial d^{[2]}_{lik} \over \partial u^j} -
{\partial d^{[2]}_{ljk} \over \partial u^i}.  \eqno{(4.7)}$$

\noindent{\bf Theorem 4.1 [9]}. Under the assumption
$\det a^{[2]}_{ij} (u) \neq 0,$ symplectic operators of the form (4.3)
exist if and only if $M$ is an even-dimensional manifold with an
almost symplectic structure.

We remind that a manifold with an almost symplectic structure is
an even-dimensional $M$ endowed with a nondegenerate skew-symmetric
metric  $g_{ij} (u)$.

Let $a^{[2]}_{ij} (u) = g_{ij} (u) $.  Consider an arbitrary
symplectic operator of the form (4.2) and introduce coefficients
$\Gamma ^i_{jk} (u)$ by the relation
$$b^{[2]}_{ijk} (u) = 2 g_{is} (u) \Gamma ^s_{jk} (u).$$

\noindent{\bf Lemma 4.2 [9]}. Coefficients $\Gamma ^i_{jk} (u)$
define a symplectic connection on $(M, g_{ij})$, i.e. a differential
geometric connection compatible
with the almost symplectic structure of the manifold:
$$\nabla _k g_{ij} \equiv {\partial g_{ij} \over \partial u^k} -
g_{is} \Gamma ^s_{jk} - g_{js} \Gamma ^s_{ik} = 0$$

It turns out that any symplectic connection $\Gamma ^i_{jk}(u)$
on a manifold with an almost symplectic
structure $(M,g_{ij})$ generates a unique symplectic operator
(4.3).

\noindent{\bf Theorem 4.2 [9]}. If $(M, g_{ij})$ is a manifold with
an almost symplectic structure then there exists a one-to-one
correspondence between the symplectic connections $\Gamma ^i_{jk}$
on $(M, g_{ij})$ and closed 2-forms of the shape (4.2), $m=2$,
on the space $L(M)$ of smooth loops on $M$. This correspondence
is expressed by the formula
$$\omega(\xi,\eta) =\int _{S^1} \bigl \{<\xi, \nabla ^2_{\nu} \eta>
+ {1 \over 2} <\nu, R(\xi,\eta)\nu> +
<\nu,\sum _{(\eta, \xi, \nu)}R (\eta,\xi)\nu>$$
$$+
<T(\nu,\xi), T(\nu,\eta)> + <T(\xi, \eta), \nabla _{\nu} \nu> \bigr \} dx
 \eqno{(4.8)}$$
where $<\xi,\eta>= g_{ij} \xi ^i \eta ^j$, the sum $\sum _{(\eta, \xi,
\nu)}$ is taken with respect to all cyclic permutations of the elements
$(\eta, \xi, \nu)$, $\ \nu$ is the velocity vector of the loop $\gamma
(x)$, i.e. $\nu ^i = u^i_x$, $\ \nabla _{\nu}$ is the covariant derivation
along $\gamma$, $\ [T(\xi,\eta)]^i = T^i_{kl} \xi
^k \eta ^l, \ T^i_{jk} \equiv \Gamma ^i_{jk} - \Gamma ^i_{kj}$, is the
torsion tensor of the connection and
$[R(\xi,\eta)\zeta] ^i = R^i_{jkl} \xi ^k \eta ^l \zeta ^j$,
$$R^i_{jkl} = - {\partial \Gamma ^i_{jl} \over \partial u^k} +
{\partial \Gamma ^i_{jk} \over \partial u^l} -
\Gamma ^i_{pk} \Gamma ^p_{jl} + \Gamma ^i_{pl} \Gamma ^p_{jk},$$
is the curvature tensor of the connection.

The formula (4.8) gives the general form of closed 2-forms of
the shape (4.2), m = 2, and also describes all the symplectic
operators of the form (4.3).

On symplectic manifolds there is a special class of symmetric
symplectic connections, satisfying the condition $T^i_{jk}=0$.
Such connections exist on $(M,g_{ij})$ if and only if the manifold
$(M,g_{ij})$ is in fact a symplectic one, i.e., $(dg)_{ijk} = 0$.

\noindent{\bf Corollary 4.1 [9]}. If a symplectic connection
$\Gamma ^i_{jk}(u)$ is a symmetric one then the closed 2-form
(4.8) has the presentation
$$\omega (\xi,\eta) = \int _{S^1} \bigl \{
<\xi, \nabla^2_{\nu} \eta> + {1 \over 2}<\nu, R(\xi,\eta) \nu>
\bigr \} dx \eqno {(4.9)}$$
This formula describes a most natural class of homogeneous
symplectic forms of the second order on loop spaces of symplectic
manifolds.

It is worth being noted that the class of symplectic connections
on manifolds with almost symplectic structures is rich enough.
Namely, any such a manifold has infinitely many distinct
symplectic connections and correspondingly infinitely
many distinct symplectic forms (4.8) on its loop space.
Similarly, any symplectic manifold has infinitely many
symmetric symplectic connections and correspondingly symplectic
forms of the shape (4.9) on its loop space.
\vskip .3 in
\noindent{\bf 5 Lagrangian systems and symplectic structures
on loop spaces}
\vskip .2 in
We shall consider in this section two-dimensional Lagrangian
systems generated by the actions of the form ([14,12])
$$S=\int \bigl ( g_i (x, u, u_x,...)u^i_t +
h(x,u,u_x,...) \bigr ) dxdt \eqno{(5.1)}$$
where $g_i,\  h$ are arbitrary functions depending on the independent
space variable $x$, some $N$ fields $u^i(x,t)$ and their
derivatives $u^i_{(n)} = \partial ^n u^i / \partial x^n $ with
respect to $x$.

The corresponding Lagrangian system
$${\delta S \over \delta u^i} = 0$$
has the following form
$$M_{ij}u^j_t = {\delta H \over \delta u^i}  \eqno{(5.2)}$$
where
$$M_{ij} ={\partial g_i \over \partial u^j_{(n)}}
{d^n \over d x^n} - (-1)^n {d^n \over d x^n} \circ
{\partial g_j \over \partial u^i_{(n)}} \eqno{(5.3)}$$
and
$$H=\int h dx$$
\noindent{\bf Lemma 5.1 [12,14]}. The differential operator $M_{ij}$ is
a symplectic operator and the bilinear form
$$\omega (\xi,\eta) = \int _{S^1} \xi^i M_{ij} \eta ^j dx, \eqno{(5.4)}$$
where $\xi, \eta \in T_{\gamma} L(M), \ $ is a symplectic form
on the loop space $L(M)$.

In other words, the inverse operator $K^{ij} = (M^{-1})^{ij}$,
such that $K^{ij} M_{jk} = \delta ^i_k$, gives always  a nonlocal
Poisson bracket
$$\{ u^i(x), u^j(y) \} = K^{ij}[u(x)] \delta (x-y) \eqno{(5.5)}$$
and hence we have always the following nonlocal Hamiltonian
representation for Lagrangian systems generated by the actions (5.1):
$$u^i_t = K^{ij} {\delta H \over \delta u^j} \equiv
\{ u^i (x), H \} \eqno{(5.6)}$$

The corresponding symplectic representation for these Lagrangian systems
 has the form ([14])
$$\omega (\delta u, u_t) = \delta H \eqno{(5.7)} $$
where the relation (5.7) is valid for arbitrary variations
$\delta u^i(x)$ of the fields $u^i(x),\ H$ is a functional on the loop
space $L(M)$.

Some special actions of the form (5.1) generate important nonlinear
equations of mathematical physics and the field theory such as
nonlinear sigma models (see above section 3), Monge-Amp\`ere
equations, some systems of hydrodynamic type, Krichever-Novikov
equation and so on and the corresponding symplectic representations
are very useful and effective for investigation of integrability and
the procedures of averaging. Of course, it is easy to generalize
this construction for arbitrary degenerate Lagrangian quasilinear
with respect to higher $t$-derivatives of the fields $u^i(x,t)$.

Let's consider some examples.

\noindent{\it Example 5.1}. The Korteweg-de Vries equation (KdV).

Consider the following simple action of the type (5.1)
$$S= \int \bigl ( {1 \over 2} u_x u_t - {1 \over 2} u^2_{xx} -
u^3_x \bigr ) dxdt$$

The symplectic representation (5.2)-(5.3) for the corresponding
Lagrangian equation $\delta S / \delta u = 0$ has the form
$$ {d \over dx} (u_t) = - {\delta H \over \delta u},$$
$$H= \int \bigl ( u^3_x + {1 \over 2} u^2_{xx} \bigr ) dx$$

After the field transformation $v(x) = u_x (x)$
we have obtained the usual Hamiltonian representation of the
KdV equation:
$$v_t = {d \over dx} {\delta H \over \delta v},$$
$$H= \int \bigl ( v^3 + {1 \over 2} v^2_x \bigr ) dx$$
$$v_t =6vv_x - v_{xxx}$$

\noindent{\it Remark 5.1}. Note that if $v^i (x) = u^i_x (x)$
then for any functional $H$ the relation
$${\delta H \over \delta u^i} = -
{d \over dx} {\delta H \over \delta v^i} $$
is valid for the variation derivatives.

\noindent{\it Example 5.2 }[10]. The Krichever-Novikov equation (KN).

Consider the special action of the form (5.1)
$$S=
\int \biggl ( {1 \over 2} {u_t \over u_x} - {1 \over 2}
{u^2_{xx} \over u^2_x} - {1 \over 3} {R(u) \over u^2_x} \biggr )
dxdt, \eqno{(5.8)}$$
where $R(u) = a_3 u^3 + a_2 u^2 + a_1 u + a_0 $ is an arbitrary
polynomial of degree three, $a_i=const,\  i= 0, 1, 2, 3$.

In this case  the Lagrangian equation $\delta S / \delta u =0$
has the symplectic representation (5.2)--(5.3) with the
symplectic operator
$$M= - {1 \over u_x} {d \over dx} \circ {1 \over u_x}
 \eqno{(5.9)}$$
$$\bigl ( {1 \over u_x} {d \over dx} {1 \over u_x} \bigr ) u_t
={ \delta H \over \delta u},  \eqno{(5.10)}$$
$$H=\int \bigl ( {1 \over 2} {u^2_{xx} \over u^2_x} +
{1 \over 3} {R(u) \over u^2_x} \bigr ) dx  \eqno{(5.11)}$$

The formulas (5.10)--(5.11) give the well-known symplectic
(and Hamiltonian) representation ([20]) for the Krichever-Novikov
equation ([21])
$$u_t = u_{xxx} - {3 \over 2} {u^2_{xx} \over u_x} + {R(u) \over
u_x},  \eqno{(5.12)}$$
which is a completely integrable equation with an infinite
collection of conservation laws, a representation of zero
curvature with parameter on an elliptic curve and so on,
and plays an important role in the contemporary theory of
integrable nonlinear systems of mathematical physics.

We recall that the equation (5.12) appeared for the first time
in [21] (in slightly other form, see also [22, 23]) as the
deformation equation (by virtue of the Kadomtsev-Petviashvili
(KP) equation) for the Tyurin parameters corresponding to
a general solution of the commutation equation of ordinary
differential operators for a spectral curve of genus $g=1$
and fixed rank $l=2$ of the vector bundle of common
eigenfunctions over the spectral curve (we note that the
corresponding integrable Boussinesq type system of deformation
of Tyurin parameters by virtue of KP equation for general solution
of the commutation equation of ordinary differential operators
of genus $g=1$ and rank $l=3$ is obtained in author's works
[24, 25, 13]).
Correspondingly any solution of the KN equation determines a
solution of KP by an explicit formula [21,22].
It is shown in [26, 27] that among
all nonlinear equations (up to local changes of the
field variable $u=u(v)$) of the Korteweg-de Vries type:
$$u_t =u_{xxx} + f(u,u_x,u_{xx}),$$
 having infinite series of local conservation laws, only the
KN equation (in general case, i.e., when the polynomial
$R(u)$ has no multiple roots) can not be reduced to the KdV
equation by differential substitutions of the form
$$u=F(v,v_x,...,v_{(m)})$$
although, generally speaking,
the question of possible connection between the KN and KdV
equations by more complicated transformations is still unsolved.

The KN equation appeared as the first nonformal example of a
system whose Hamiltonian structure is defined by a nontrivial
differential symplectic operator.
The action (5.8) gives Lagrangian representation for the KN
equation ([10]) and explains natural origin of the symplectic
operator (5.9).

The KN equation (5.12) is related with the Hamiltonian
equation ([10])
$$v_t = - \bigl ( {d \over dz} \bigr ) ^{-1} {\delta H
\over \delta v}, \eqno{(5.13)}$$
$$H = - \int \bigl [ {1 \over 2} {v_{zz}^2 \over v^3_z} +
{1 \over 3} R(z) v_z^3 \bigr ] dz, \eqno{(5.14)}$$
by the point transformation of hodograph type
$$\cases { x=v, &\ \cr u=z & \ \cr} \eqno{(5.15)}$$
One can assume that $v(z)$ is a new scalar field which is
the inverse of the field $u(x)$, i.e., $u(v(z)) =z $ and
$v(u(x))=x.$

Introducing the new field $w(z) =v_z(z)$ we obtain a canonical
Hamiltonian representation of the Krichever-Novikov equation
(5.12) ([10]):
$$w_t = {d \over dz} {\delta H \over \delta w}, \eqno{(5.16)}$$
$$H= - \int \bigl [ {1 \over 2} {w^2_z \over w^3} +
{1 \over 3} R(z) w^3 \bigr ] dz. \eqno{(5.17)}$$

Note that the Hamiltonian $H$ is not invariant with respect to
translations in $z$.

As it was shown by author in [10] the KN equation is simply
connected also with the canonical Hamiltonian system
$$\cases {q_t = {\delta F \over \delta p},&\ \cr
p_t = - {\delta F \over \delta q},&\ \cr} \eqno{(5.18)}$$
where the Hamiltonian $F[p, q]$ in conjugate variables
$p(z),\  q(z)$ has the form
$$F[p,q] =$$
$$ \int \biggl ( p(z) \biggl ({q_{zzz} \over q^3_z} -
{3 \over 2} {q^2_{zz} \over q^4_z} - R(z) q^2_z \biggr )
+ {1 \over 3} R(z) q^3_z - {3 \over 2} {q^2_{zz} \over q^3_z}
\biggr ) dz \eqno{(5.19)}$$

  In [28, 29, 30] symplectic
structures of the degenerate
KN equation ($R(u) \equiv 0$) are considered
(as it is known ([27]), in this case and always, when
the polynomial $R(u)$ has multiple roots, the KN equation reduces to
the KdV equation by differential substitution). The degenerate
KN equation
has the form
$$u_t = u_x S[u],  \eqno{(5.20)}  $$
where $$S[u] = {u_{xxx} \over u_x} - {3 \over 2} {u^2_{xx} \over
u^2_x}$$
is the Schwarzian of $u$, and has a compatible pair of
differential symplectic operators found by Dorfman ([28--30, 12])
and a number of other interesting properties studied by various
authors (see [31], for example). For multicomponent analog of the
degenerate KN equation (5.20) multisymplectic representation is
found by Fordy and Antonowicz [32]. In [63] a (2+1)-dimensional version
of the degenerate KN equation (5.20) was studied by Dorfman and Nijhoff.

It is interesting that as it was discovered by author and Ferapontov
([15], see also [16-17] and sections 7 and 9 in this paper)
the symplectic operator (5.9)
and its multicomponent generalization
$$M_{ij} = {1 \over K} {1 \over u^i_x} {d \over dx} \circ
{1 \over u^j_x}, \eqno{(5.21)}$$
$K= const$,
arises
naturally in the Dubrovin-Novikov theory of Hamiltonian systems of
hydrodynamic type developed in [1-4].
The Hamiltonian structure of these systems introduced and studied
by Dubrovin and Novikov in [2] is generated by metrics of zero
curvature (all these Hamiltonian systems of hydrodynamic type
have a natural
Lagrangian representation of the considered above type (5.1)--(5.3)
and we shall consider them in the section 7).
It was shown in [15] that with the help of the operator
(5.21) one can adequately extend the theory of Hamiltonian
systems of hydrodynamic type to metrics of arbitrary constant
Riemannian curvature $K$. In particular, such Hamiltonian diagonalizable
systems of hydrodynamic type will be integrable (they are
semi-Hamiltonian in the sense of Tsarev [5--6]). All systems of
hydrodynamic type derived by averaging of well-known integrable
equations of the soliton theory (KdV, Nonlinear Schr\"odinger,
Sine-Gordon and so on) have Hamiltonian structures generated by
metrics of non-zero constant Riemanian curvature (along with
structures of Dubrovin-Novikov type) (see [19]).

\vskip .3 in
\noindent{\bf 6 Symplectic and Poisson structures
of the Monge-Amp\`ere equations}
\vskip .2 in
\noindent{\it Example 6.1}. The Monge-Amp\`ere equations.

Consider the following action of type (5.1) [50]
$$S = \int [{1 \over 2} u_x^2 q_t -u_x q_x u_t -
{1 \over 2} q^2 u_{xx} + \Phi (x,t,u,u_x)] dx dt, \eqno{(6.1)}$$
where $\Phi (x,t,u,u_x)$ is an arbitrary function.

The corresponding Lagrangian system $\delta S / \delta u^i =0$
has natural symplectic representation (5.2)-(5.3) of the form
$$M {\pmatrix {u\cr q \cr}}_t={\pmatrix {\delta F / \delta u \cr
\delta F / \delta q \cr}}, \eqno{(6.2)}$$
where
$$M= {\pmatrix {q_x {d \over dx} + {d \over dx} \circ q_x & -u_{xx} \cr
u_{xx} & 0 \cr}} \eqno{(6.3)}$$
is a symplectic operator,
$$F= \int [{1 \over 2} q^2 u_{xx} - \Phi (x,t,u,u_x)]dx. \eqno{(6.4)}$$
The symplectic system (6.2)-(6.4) is equivalent to the Monge-Amp\`ere
equation ([50])
$$u_{xx} u_{tt} - (u_{xt})^2 = {\delta \Phi \over \delta u}.
\eqno{(6.5)}$$

In this case of the Monge-Amp\`ere equations the Poisson bracket
(5.5) is also local and corresponding Hamiltonian operator
$K^{ij} = (M^{-1})^{ij}$ has the form ([50])
$$K={\pmatrix {0& {1 \over u_{xx}}\cr
-{1 \over u_{xx}} & {q_x \over u_{xx}^2}{d \over dx} +
{d \over dx} \circ {q_x \over u_{xx}^2}\cr}} \eqno{(6.6)}$$

And what is more this fact is valid for all local Hamiltonian
operators of the form
$$K={\pmatrix {0&B\cr -B& A{d \over dx} +A \circ {d \over dx}\cr}},
\eqno{(6.7)}$$
where $B$ and $A$ are some functions of $x,u,u_x,u_{xx},...,q,q_x,
q_{xx},...\ $ ($B \neq 0$), which were considered by Nutku and
Sar$\imath$o$\breve{g}$lu in [50] in connection with Poisson
structures of the Monge-Amp\`ere equations.
Namely, for any local Hamiltonian differential operator of
the form (6.7)
the corresponding inverse symplectic operator $M$ is also
local differential operator and has the form
$$M={\pmatrix {{A \over B^2} {d \over dx} + {d \over dx} \circ
{A \over B^2} & - {1 \over B} \cr {1 \over B} & 0\cr}} \eqno{(6.8)}$$

\noindent{\bf Theorem 6.1}. The operator
$$M={\pmatrix {\bigl [ {\partial g \over \partial u_x} -
\bigl ( {\partial g \over \partial u_{xx}} \bigr )_x \bigr ] {d
\over dx} + {d \over dx} \circ \bigl [ {\partial g \over
\partial u_x} - \bigl ( {\partial g \over \partial u_{xx}} \bigr )_x
\bigr ] & {\partial g \over \partial q}\cr -{\partial g \over
\partial q}&0\cr}} \eqno{(6.9)}$$
is a symplectic operator of the form (6.8) for arbitrary function
$g(x,t,u,u_x,u_{xx},q)$.
The corresponding local Hamiltonian operator $K$ has the form
(6.7) where
$$B= - {1 \over  (\partial g / \partial q)},
\ \ \ A=
{{ {\partial g \over \partial u_x} -
\bigl ( {\partial g \over \partial u_{xx}} \bigr )_x}
\over ( \partial g / \partial q)^2} \eqno{(6.10)}$$
\vskip .3 in
\noindent{\bf 7 Symplectic and nonlocal Poisson structures
of homogeneous and nonhomogeneous systems of hydrodynamic type}
\vskip .2 in
\noindent{\it Example 7.1}. Systems of hydrodynamic type.

Consider the following action of type (5.1)
$$S= \int \bigl ( {1 \over 2} g_{ij} u^i_x u^j_t -
h(u_x) \bigr ) dxdt \eqno{(7.1)}$$
where $(g_{ij})$ is a constant nondegenerate symmetric tensor,
$h(v)$ is an arbitrary function.

The corresponding Lagrangian system $\delta S / \delta u^i = 0$
has the symplectic representation (5.2)-(5.3) of the form
$$g_{ij} {d \over dx} (u^j_t) = - {\delta H \over \delta u^i},
\eqno{(7.2)}$$
$$H= \int h(u_x) dx$$

After the field transformation $ v^i (x) = u^i_x (x)$
(see Remark 5.1) we obtain a Hamiltonian system of hydrodynamic type
in flat coordinates
$$v^i_t = g^{ij} {d \over dx} {\delta H \over \delta v^j}, \eqno{(7.3)}$$
$$H = \int h(v) dx$$

In other words, the action (7.1) gives the general class of
Hamiltonian systems of hydrodynamic type (written in flat coordinates
$(v^1,...,v^N)$), introduced and studied by Dubrovin and Novikov [1-4].

We remind very briefly the basic necessary for us notions and results
on Poisson structures of hydrodynamic type.

Consider the one-dimensional systems of hydrodynamic type, i.e.,
in other words, the one-dimensional evolution quasilinear systems
of the first order
$$u^i_t=v^i_j (u) u^j_x, \eqno{(7.4)}$$
where $v^i_j (u) $ is an arbitrary $N \times N$ matrix function of
$u=(u^1,...,u^N),\ u^i = u^i (x,t),\ i=1,...,N.$

The Hamiltonian systems of hydrodynamic type considered by
Dubrovin and Novikov in [1-4] have the form
$$u^i_t =\{ u^i,H \}, \eqno{(7.5)}$$
where $H$ is a functional of hydrodynamic type, i.e.
$$H=\int h(u) dx, \eqno{(7.6)}$$
and the Poisson bracket has the form
$$\{ u^i (x), u^j (y) \} =
g^{ij} (u(x)) \delta _x (x-y) + b^{ij}_k (u(x)) u^k_x \delta (x-y)
\eqno{(7.7)}$$
(the Poisson bracket of Dubrovin-Novikov type [2]). It was shown
in [2] that if $\det [g^{ij} (u)] \neq 0$ then the expression (7.7)
gives a Poisson bracket if and only if

\item{(1)} $g^{ij} (u)$ is a metric of zero Riemannian curvature
(that is, simply a flat metric),

\item{(2)} $b^{ij}_k (u) = - g^{is} (u) \Gamma ^j_{sk} (u),$
where $\Gamma^j_{sk} (u) $ are the coefficients of the differential
geometric connection generated by the metric $g^{ij},$
that is, the only symmetric connection compatible with the
metric (the Levi-Civita connection).

Thus, there exist always local variables $v^i= v^i(u)$ in which
the Poisson bracket (7.7) is simply constant:
$$\{ v^i(x),v^j(y) \} =\varepsilon ^i \delta ^{ij} \delta _x
 (x-y), \eqno{(7.8)}$$
where $\varepsilon ^i = \pm 1,\ i=1,...,N.$
Correspondingly, it is easy to give a Lagrangian description
(7.1) to these systems. The considered Hamiltonian systems of
hydrodynamic type have the form
$$u^i_t = [\nabla ^i \nabla _j h(u)] u^j_x, \eqno{(7.9)}$$
where $\nabla$ is the covariant derivative generated by a zero
curvature metric. As it was shown in [5,6], if the
Hamiltonian system of hydrodynamic type (7.9) has the
Riemann invariants (that is, the matrix $v^i_j (u)=
\nabla ^i \nabla _j h(u)$ is diagonalizable), then it is
integrable.

Multidimensional Poisson structures of hydrodynamic type
were introduced and studied in [3,7].

In [15] (see also [16,17]) there was proposed a nonlocal generalization of
Hamiltonian theory of the systems of hydrodynamic type (7.4)
connected with the nonlocal Poisson brackets of the form
$$ \{ u^i (x), u^j (y) \}
=$$

$$g^{ij} (u(x)) \delta _x (x - y) + b^{ij}_k (u(x)) u^k_x \delta (x-y)
+ K u^i_x (d/dx)^{-1} u^j_x \delta (x-y). \eqno{(7.10)}$$

It is simple to show that for any Hamiltonian functional $H$ of
hydrodynamic type ($H=\int h(u)dx$) the Poisson bracket of the
form (7.10) also generates always a system of hydrodynamic type
(7.4).

\noindent{\it Remark 7.1} [16]. The expression (7.10) is the
most general form of Poisson brackets with the property
to generate a system of hydrodynamic type (7.4) for any
Hamiltonian $H$ of hydrodynamic type (7.6).

\noindent{\bf Theorem 7.1 [15]} If $\det [g^{ij} (u)] \neq 0$
then expression (7.10) gives a Poisson bracket if and only
if

\item{(1)} $g^{ij} (u)$ is a metric of constant Riemannian curvature
$K$,

\item{(2)} $b^{ij}_k (u) = - g^{is} \Gamma ^j_{sk} (u),$
where $\Gamma ^j_{sk} (u)$ are the coefficients of the connection
generated by the metric $g^{ij}$ (the Levi-Civita connection).

The corresponding Hamiltonian systems of hydrodynamic type
have the form
$$u^i_t = [\nabla ^i \nabla _j h(u) + K \delta ^i_j h(u)] u^j_x,
\eqno{(7.11)}$$
where $\nabla$ is the covariant derivative generated by a metric
of constant Riemannian curvature $K$.

If $\det [g^{ij}(u)] =0$ then description of the nonlocal Poisson
brackets (7.10) much more complicated [16].

Note that the Poisson brackets (7.7) and (7.10)
define natural differential geometric
Poisson brackets on loop spaces of arbitrary flat
 and constant Riemannian curvature manifolds correspondingly
by analogy with symplectic structures of differential geometric
type  on loop spaces of arbitrary Riemannian manifolds [8,9]
(see also (3.4) and the sections 2--4 in this paper).

Consider now nonhomogeneous systems of hydrodynamic type
$$u^i_t = v^i_j (u) u^j_x + f^i (u). \eqno{(7.12)}$$

Local Poisson structures for the systems (7.12)
studied by Dubrovin and Novikov [3,4]
have also a natural nonlocal generalization.
We shall consider here nonlocal Hamiltonian nonhomogeneous
systems of hydrodynamic type
$$u^i_t = \{u^i, H \}, \eqno{(7.13)}$$
where $H$ is a functional of hydrodynamic type (7.6), and
the nonlocal nonhomogeneous Poisson bracket of
hydrodynamic type has the form
$$\{ u^i (x), u^j (y) \} = g^{ij} (u(x)) \delta _x (x-y) +
b^{ij}_k (u(x)) u^k_x \delta (x-y) +$$

$$K u^i_x (d/dx)^{-1} u^j_x \delta (x-y) + \omega ^{ij} (u(x)) \delta (x-y)
\eqno{(7.14)}$$
It is easy to show that the expression (7.14) is a Poisson bracket
if and only if it is a sum of two compatible Poisson brackets
(7.10) and (1.3).

\noindent{\bf Theorem 7.2 [52]}. The Poisson brackets (7.10) and
(1.3) are compatible if and only if the bivector $\omega ^{ij} (u)$
is a Killing bivector on the manifold $(M, g^{ij})$ of constant
Riemannian curvature $K$, that is,
$$\nabla ^i \omega ^{jk} + \nabla ^j \omega ^{ik} = 0, \eqno{(7.15)}$$
where $\nabla$ is the covariant derivative generated by the metric
$g^{ij}(u)$ of constant Riemannian curvature $K$.

\vskip .3 in
\noindent{\bf 8 On some integrable nonhomogeneous systems of
hydrodynamic type}
\vskip .2 in
In this section we shall consider
some special class of nonhomogeneous
systems of hydrodynamic type with quadratic nonlinearity ([49,48]):

$$u^i_t = a^i u^i_x + \sum _{k,j} b^i_{jk} u^j u^k +
\sum _k c^i_k u^k , \eqno{(8.1)} $$
where a summation over repeating indices is not assumed in this section,
$a^i, b^i_{jk}$ and $c^i_k$ are constant tensors, $i,j,k= 1,...,N$.

There are a number of well-known integrable systems among (8.1).

\noindent{\it Example 8.1}. The N-wave equation.

For example, integrable real-valued exact resonance
system of parametric interaction of three wave packets
in nonlinear optics ($N=3$)
is a special case of (8.1):
$$\cases{u^1_t=a^1 u^1_x - \varepsilon u^2 u^3,&\ \cr
u^2_t=a^2 u^2_x + \varepsilon u^1 u^3, &\ \cr
u^3_t=a^3 u^3_x + \varepsilon u^1 u^2, &\ \cr}
\eqno{(8.2)}$$
where $a^i, \ \varepsilon $ are some constants.

\noindent{\it Example 8.2}. The KdV equation.

Consider the KdV equation (see also Example 5.1) as evolution system
with respect to $x$:
$$\cases {u^1_x = u^2,&\ \cr
u^2_x=u^3,&\ \cr
u^3_x=-u^1_t + 6u^1 u^2 &\ \cr} \eqno{(8.3)}$$

It was shown in [47] that the KdV system (8.3) is Hamiltonian with
respect to some nonhomogeneous Poisson structures of
hydrodynamic type (7.14) which are
in fact induced by well-known Magri and Gardner brackets for KdV.
After the local quadratic unimodular change of field variables [48]

$$\cases {u^1 =(w^1-w^3)/ \sqrt {2}, &\ \cr
u^2=w^2, &\ \cr
u^3= (w^1+w^3)/ \sqrt {2} + (w^1 -w^3)^2 &\ \cr} \eqno{(8.4)}$$
we obtain Hamiltonian representation for the KdV system (8.3) ([48]),
generated by the simplest infinite dimensional Kac-Moody Lie algebra
$\widehat {sl} (2)$ and some quadratic Hamiltonian $H$:
$$w^i_x = M^{ij} (t) {\delta H \over \delta w^j}, \eqno{(8.5)}$$
$$H= - \int [(w^1 -w^3)^2 - \sqrt {2} (w^1 + w^3)]dt, $$
$$(M^{ij} (t)) =
{\pmatrix { 1 & 0 & 0 \cr
            0 & -1& 0 \cr
            0 & 0 & -1 \cr}} {d \over dt} +
{\pmatrix {0 & -2w^3(t) & 2 w^2(t) \cr
2 w^3 (t) &  0 & 2w^1 (t) \cr
-2w^2(t) & -2w^1(t) & 0 \cr }}, \eqno{(8.6)}$$
where $M$ is the Hamiltonian operator given by $\widehat {sl} (2)$ [48].

The second nonhomogeneous Poisson structure of hydrodynamic type
for the KdV system (8.3) is compatible with (8.6) and has the
following form [48]:
$$(L^{ij} (t)) = {1 \over 2} {\pmatrix {1 & 0 & 1 \cr
0&0&0\cr
1&0&1\cr}} {d \over dt} +$$
$$ (w^1 - w^3) {\pmatrix {0&1&0\cr
-1&0&-1\cr
0&1&0\cr}} + {1 \over \sqrt {2}} {\pmatrix {0&1&0\cr
-1&0&1\cr
0&-1&0\cr}}, \eqno{(8.7)}$$
here the metric $g^{ij} (w)$ in (8.7) is degenerate (see (7.14)).

Correspondingly, second Hamiltonian representation for the KdV system (8.3)
has the form

$$w^i_x = L^{ij} (t) {\delta H \over \delta w^j},$$
where $H$ is the quadratic Hamiltonian
$$H= - {1 \over 2} \int [(w^1)^2 -(w^2)^2 - (w^3)^2]dt $$

The second Poisson structure (8.7) for the KdV system (8.3)
is generated by the three-dimensional nilpotent non-Abelian
Lie algebra $\cal {G} _0$ (it is the Lie algebra of type II
according to the Bianchi classification of three-dimensional
Lie algebras) and 2-cocycle on its loop algebra.
Note that any nilpotent non-Abelian Lie algebra contains
subalgebra isomorphic to $\cal {G} _0$.

Consider the special system of type (8.1):
$$u^i_t = a^i u^i_x + u^i \sum _k (a^i - a^k)u^k,
 \eqno{(8.8)}$$
where $a^i \neq a^j, $ if $\ i \neq j,\  i,j= 1,...,N$.

\noindent{\bf Theorem 8.1}. The system (8.8) is integrable and
equivalent to some integrable homogeneous diagonal Hamiltonian
system of hydrodynamic type by a combination of a
reciprocal transformation and some changes of the fields and
independent variables.

First of all, let us introduce new fields variables $w^i(x)$
by the relation
$$u^i = e^{w^i}$$

Then our system (8.8) will have the following form:
$$w^i_t = a^i w^i_x + \sum _k(a^i - a^k) e^{w^k} \eqno{(8.9)}$$

Consider so-called reciprocal transformations of the system (8.9),
or, in other words, transformations of the independent
variables $x$ and $t$ in according to a solution
$w(x,t)$ of the system (see also [42,51]):

$$dx' = \varphi _1 (x,t,w) dx - \psi _1 (x,t,w) dt,$$
$$dt' = \varphi _2 (x,t,w) dx - \psi _2 (x,t,w) dt, \eqno{(8.10)}$$
where
$${\partial \varphi _i (x,t,w) \over \partial t} +
{\partial \psi _i (x,t,w) \over \partial x} = 0,\ \ i=1,2, \eqno{(8.11)}$$
$$\Delta = \varphi_1 \psi_2 - \varphi _2 \psi _1 \neq 0.$$

Here $w^i (x,t)$ is an arbitrary solution of the system (8.9),
$\varphi _i,\ \psi _i,\ i=1,2,$ are generated by
some conservation laws (8.11) of the system (8.9).

Let us consider the following two conservation laws of the system (8.9):
\item{1)} $\varphi _1 = - \sum _k e^{w^k}, \ \psi _1 = \sum _k a^k
e^{w^k}$

\item{2)} $\varphi _2 =0,\ \psi _2 =1$

Using the relations
$$w^i_x = w^i_{x'} \varphi _1 + w^i_{t'} \varphi _2,$$
$$w^i_t = -w^i_{x'} \psi _1 - w^i_{t'} \psi _2,$$
after corresponding reciprocal  transformation (8.10)-(8.11)
we obtain the following nonhomogeneous system of hydrodynamic
type:
$$w^i_{t'} = \bigl ( \sum _k (a^i - a^k) e^{w^k} \bigr )
(w^i_{x'} -1) \eqno{(8.12)}$$
After transformations
$$v^i = w^i - x' $$
and
$$x'' = - e^{-x'}$$
we obtain diagonal weakly nonlinear semi-Hamiltonian homogeneous
system of hydrodynamic type which is integrable due to the Tsarev theorem
[5,6]:

$$v^i_{t'} = \bigl ( \sum _k (a^i - a^k) e^{v^k} \bigr ) v^i_{x''}
\eqno{(8.13)}$$

\vskip .3 in
\noindent{\bf 9 Killing-Poisson bivectors on Riemannian
manifolds and integrable bihamiltonian hierarchies of
Heisenberg magnet type}
\vskip .2 in

The Killing-Poisson bivector $\omega ^{ij}(u)$ on $N$-dimensional
Riemannian manifold $(M, g_{ij})$ is by definition (see [53])
a skew-symmetric tensor on $M\ $ ($\omega ^{ij} = - \omega ^{ji}$)
which satisfies two the well-known relations: (1.1) (the Jacobi
identity for the Poisson bivector $\omega^{ij}(u)$) and (7.15)
(the identity for the Killing bivector $\omega ^{ij}(u)$
on the Riemannian manifold $(M, g_{ij})$ with the Levi-Civita
connection $\nabla$ generated by the metric $g_{ij}$).
It was shown in section 7 (see also [52])
that Killing-Poisson bivectors on the manifolds of constant
Riemannian curvature $K$ define natural nonlocal (if $K \neq 0$)
Poisson structures for
nonhomogeneous systems of hydrodynamic type.
Here we shall give complete description of the Killing-Poisson
bivectors on the manifolds of constant Riemannian curvature
in terms of Lie algebras with invariant scalar products and
show that the compatible pairs of Poisson structures given
by these bivectors generate also integrable bihamiltonian hierarchies
of Heisenberg magnet type [52].

It is easy to describe all Killing-Poisson bivectors on flat spaces
or, in other words, spaces of zero Riemannian curvature.
In this case we shall consider flat coordinates $(v^1,...,v^N)$
in which the metric
$g_{ij}$ is constant.
A tensor $\omega ^{ij} (v)$ is a Killing-Poisson
bivector on flat space $(M, g_{ij})$ if and only if
in flat coordinates $(v^1,...,v^N)$ the following conditions
are valid:
$$\omega ^{ij} (v) = c^{ij}_k v^k + d^{ij}, \eqno{(9.1)}$$
where $c^{ij}_k,\ d^{ij}$ are constants such that

\item{1)} $c^{ij}_k$ are structural constants of a Lie algebra
with invariant scalar product $<\ ,\ > $ given by
the metric $g_{ij}$: $<ad X (Y), Z> = - <Y, ad X(Z)>$;

\item{2)} $d^{ij}$ is a 2-cocycle on this Lie algebra:
$$d^{ij} = -d^{ji},\ \ \sum _{(i,j,k)} c^{ij}_s d^{sk} =0, $$
where the sum is taken with respect to all cyclic permutations of
the indices $i,j,k$.

For example, any semi-simple Lie algebra gives a Killing-Poisson
bivector on flat space (we must take the Killing metric on
the Lie algebra as a metric for corresponding flat space).
This description coincides with classification of local
nonhomogeneous Poisson brackets of hydrodynamic type [3,4].

Let us consider arbitrary metric $g^{ij}$ of constant Riemannian curvature
$K$ in canonical variables $u^1,...,u^N$:
$$(g^{ij}(u))= (\lambda (u))^2 {\pmatrix {\varepsilon _1&0&\ldots&0\cr
0& \varepsilon _2 & \ldots &0\cr
\vdots&\vdots&\ddots&\vdots\cr
0&0&\ldots&\varepsilon _N\cr}}, $$
$$\varepsilon _i = \pm 1,\ \ \lambda (u) = 1 +  {K \over 4}
\sum _i \varepsilon _i (u^i)^2$$

\noindent{\bf Theorem 9.1}. For $N=2$ a tensor $\omega ^{ij} (u)$
is a Killing-Poisson bivector on some space $(M, g_{ij})$
of constant Riemannian
curvature $K$ if and only if
in canonical variables $(u^1,u^2)$
$$(\omega ^{ij} (u))= c \lambda ^2 (u) {\pmatrix {0&1\cr
-1&0\cr}}, \eqno{(9.2)}$$
where $c$ is an arbitrary constant.

As it was shown in [52] (see also section 7)
any Killing-Poisson bivector
$\omega ^{ij} (u)$ on
the space $(M, g_{ij})$ of constant Riemannian curvature $K$
define pair of compatible Hamiltonian operators
$$M^{ij}_1 = g^{ij} (u) {d \over dx} - g^{is} (u) \Gamma ^j_{sk}
(u) u^k_x + K u^i_x \bigl ( {d \over dx} \bigr )^{-1} u^j_x,
\eqno{(9.3)}$$
$$M^{ij}_2 = \omega ^{ij}(u), \eqno{(9.4)}$$
which generate the integrable hierarchy of the generalized
$N$-component Heisenberg magnet equations
$$S_t = [S, S_{xx}],\ \ \ \ S^2 =1,$$
where $S$ is an $(N+1)$-vector and $[\ ,\ ]$ is the commutator
in an $(N+1)$-dimensional Lie algebra equipped with an invariant
inner product. The classic Heisenberg magnet corresponds to
the simplest case of two-dimensional sphere ($N=2$).
For two-dimensional sphere ${S^1}^2 +{S^2}^2 +{S^3}^2 =1 $
in coordinates of stereographic projection
$$S^1=u^1/P,\ \ \  S^2=u^2/P,\ \ \  S^3 = (P-1)/P,$$
where $P=({u^1}^2+{u^2}^2 +1)/2,$
the metric  has the form
$$(g_{ij})= {1 \over P^2}{\pmatrix {1&0\cr
0&1\cr}}.$$
The corresponding nonlocal Hamiltonian operator (9.3)
generated by the metric has the form
$$M_1 = P^2 {\pmatrix {d&0\cr 0&d\cr}} +
P{\pmatrix {u^1 u^1_x +u^2 u^2_x & u^1 u^2_x - u^2 u^1_x \cr
u^2 u^1_x - u^1 u^2_x & u^1 u^1_x +u^2 u^2_x \cr}} +
{\pmatrix {u^1_x d^{-1} \circ u^1_x & u^1_x d^{-1} \circ
u^2_x \cr u^2_x d^{-1} \circ u^1_x & u^2_x d^{-1} \circ
u^2_x \cr}} \eqno{(9.5)}$$
In according to the Theorem 9.1 the unique (up to a constant
factor) Killing-Poisson bivector on two-dimensional
sphere in these coordinates has the form
$$M_2 = (\omega ^{ij} (u)) =
{\pmatrix {0&- P^2 \cr
P^2 &0 \cr}} \eqno{(9.6)}$$
Following to the well-known general construction of bihamiltonian
equations (see [37,61,62])
consider the recursion operator $R = M_1 (M_2)^{-1}$
corresponding to the compatible Hamiltonian pair (9.5),(9.6)
and apply it to the simplest translation flow
$${\pmatrix {u^1\cr u^2\cr}}_t =  {\pmatrix {u^1\cr u^2\cr}}_x $$
We obtain the new system
$${\pmatrix {u^1\cr u^2\cr}}_t =R {\pmatrix {u^1\cr u^2\cr}}_x,$$
or
$$\cases { u^1_t =u^2_{xx} + (u^2 (u^1_x)^2 - 2 u^1 u^1_x u^2_x -
u^2 (u^2_x)^2 )/P,&\ \cr
u^2_t = -u^1_{xx} -(u^1 (u^2_x)^2 - 2u^2 u^1_x u^2_x - u^1 (u^1_x)^2)
/P &\ \cr} \eqno{(9.7)}$$
which exactly coincides with the classic Heisenberg magnet
equations
$$\vec {S}_t = \vec {S} \times \vec {S}_{xx}, \ \ \ \vec {S}^2 =1.$$

\noindent{\bf Theorem 9.2 [52]}. The compatible Hamiltonian pair
(9.5),(9.6) generates the hierarchy of the Heisenberg magnet equations.

Explicit bihamiltonian representation of the Heisenberg magnet equations
(9.7) has the form
$${\pmatrix {u^1\cr u^2\cr}}_t = M_1 {\pmatrix {\delta G / \delta u^1\cr
\delta G / \delta u^2 \cr}} = M_2 {\pmatrix {\delta H / \delta u^1\cr
\delta H / \delta u^2 \cr}} \eqno{(9.8)}  $$
with Hamiltonians
$$G= \int {{u^2 u^1_x -u^1 u^2_x} \over (2P-1)P} dx, \ \ \ \
\ \ H={1 \over 2} \int {{(u^1_x)^2 +(u^2_x)^2} \over P^2} dx \eqno{(9.9)}$$

Note that bihamiltonian representation of Heisenberg magnet equations
was found in [59,60] (in coordinates $S^1, S^2, S^3$).

Let us give a complete description of the Killing-Poisson bivectors
on $N$-dimensional sphere [52]
$$\sum _{k=1}^{N+1} (S^k)^2 =1. \eqno{(9.10)}$$
Let $c^{ij}_k$ be structural constants of some $(N+1)$-dimensional
Lie algebra equipped with invariant scalar (Euclidean) product such that
$c^{ij}_k + c^{kj}_i =0$.
Consider the Lie-Poisson bivector $\Omega ^{ij} =
c^{ij}_k S^k$
and restrict it on the sphere (9.10).
After restriction we obtain a Killing-Poisson bivector on
$N$-dimensional sphere. The converse is also true.

\noindent{\bf Theorem 9.3 [52]}. All Killing-Poisson bivectors
on $N$-dimensional sphere can be obtained by the above construction
from any Lie algebra with an invariant scalar product.

In coordinates $(u^1,...,u^N)$ of the stereographic projection
$$S^1=u^1/P,...,S^N=u^N/P,\ S^{N+1} = (P-1)/P,\ \ P=(\sum _{s=1}^N
(u^s)^2 +1) /2 \eqno{(9.11)}$$
we have
$$\omega ^{ij} = \Omega |_{S^N} = P \sum _{s=1}^N
(c^{si}_{N+1} u^s u^j - c^{sj}_{N+1} u^s u^i + c^{ij}_s u^s) +
c^{ij}_{N+1} (P-1)P \eqno{(9.12)}$$
For example, the Killing-Poisson bivector (9.6) on two-dimensional
sphere is a result of restriction of the Lie-Poisson bivector
$$\Omega = {\pmatrix {0 & S^3 & -S^2\cr
-S^3 & 0 & S^1\cr
S^2 & -S^1 & 0\cr}}.$$

In order to obtain a complete description of
the Killing-Poisson bivectors on arbitrary
spaces of constant Riemannian curvature it is
necessary to consider Lie algebras equipped with
arbitrary invariant scalar products (with arbitrary signatures)
and apply the same construction.
The corresponding Killing-Poisson bivector
gives a compatible Hamiltonian pair (9.3),(9.4) which generates
a bihamiltonian integrable $N$-component system
of Heisenberg magnet type.

\vskip .3 in
\noindent{\bf 10 Nonlinear partial differential equations of
associativity in 2D topological field theories and
nondiagonalizable integrable systems of hydrodynamic type}
\vskip .2 in

In this section we shall consider so-called nonlinear partial
differential equations of associativity in 2D topological
field theories (see [54-57]) and give their description as
integrable nondiagonalizable weakly nonlinear systems of
hydrodynamic type. For systems of this type corresponding
general differential geometric theory of integrability
connected with Poisson structures of hydrodynamic type
can be developed.

We remind very briefly following to Dubrovin [54]
the basic mathematical concepts connected with the
Witten-Dijkgraaf-E.Verlinde-H.Verlinde (WDVV) system
arising originally in two-dimensional topological
field theories [56,57] and its relations with the
Dubrovin type equations of associativity.

Consider a function $F(t),\  t=(t^1,...,t^N)$ such that
the following three conditions are satisfied for its
third derivatives denoted as
$$c_{\alpha \beta \gamma} (t) = {\partial ^3 F(t) \over
\partial t^{\alpha} \partial t^{\beta} \partial t^{\gamma}}:$$

\item{1)} normalization, i.e.,
$$\eta _{\alpha \beta} = c_{1 \alpha \beta} (t)$$
is a constant nondegenerate matrix;

\item{2)} associativity, i.e., the functions
$$c^{\gamma}_{\alpha \beta} (t) = \eta ^{\gamma \epsilon}
c_{\epsilon \alpha \beta} (t)$$
for any $t$ define a structure of an associative algebra
$A_t$ in the N-dimensional space with a basis $e_1,...,e_N$:
$$e_{\alpha} \cdot e_{\beta} = c^{\gamma}_{\alpha \beta} (t) e_{\gamma}$$

\item{3)} $F(t)$ must be quasihomogeneous function of its variables:
$$F(c^{d_1} t^1,...,c^{d_N} t^N) = c^{d_F} F(t^1,...,t^N)$$
for any nonzero $c$ and for some numbers $d_1,...,d_N, d_F$.

The resulting system of equations for $F(t)$ is called the
Witten-Dijkgraaf-E.Verlinde-H.Verlinde (WDVV) system [56,57] (see
also [54-55]).
It was shown by Dubrovin [54] that solutions of the WDVV system
can be reduced by a linear change of coordinates to two special
types:

\item{(1)} in the most important physically case
$$F(t) = {1 \over 2} (t^1)^2 t^N + {1 \over 2} t^1
\sum ^{N-1}_{\alpha =2} t^{\alpha} t^{N - \alpha +1} +
f(t^2,...,t^N) \eqno{(10.1)}$$
for some function $f(t^2,...,t^N)$

\item{(2)} in some special case
$$F(t) = {c \over 6}(t^1)^3 + {1 \over 2} t^1
\sum ^{N-1}_{\alpha =1} t^{\alpha  }
t^{N- \alpha +1} + f(t^2,...,t^N) \eqno{(10.2)}$$
for a nonzero constant $c$.

If $N=3$ (first nontrivial case for the condition of
associativity in algebra $A_t$)
then for the first type solutions (10.1) of the WDVV system
the associativity condition in algebra $A_t$ gives the
following nonlinear partial differential equation for the
function $f(x,y)$ [54]:
$$f^2_{xxy} = f_{yyy} + f_{xxx} f_{xyy} \eqno{(10.3)}$$

Let us introduce new variables $a,b,c,d$ such that
$$a = f_{xxx},\ \ b=f_{xxy},$$
$$c = f_{xyy},\ \ d=f_{yyy}\eqno{(10.4)}$$

Conditions of compatibility have the form:
$$\cases {a_y = b_x,&\ \cr
b_y = c_x,&\ \cr
c_y = d_x &\ \cr}  \eqno{(10.5)}$$

Besides, we have the relation
$$d=b^2 - ac \eqno{(10.6)}$$
from (10.3).

So, the equation (10.3) is equivalent to the
following homogeneous system of hydrodynamic type
$${\pmatrix {a \cr
b \cr
c \cr}}_y ={\pmatrix {0 & 1 & 0 \cr
                      0 & 0 & 1 \cr
          - c & 2b & - a \cr}} {\pmatrix {a \cr
b \cr
c \cr}}_x \eqno{(10.7)}$$

The system (10.7) is nondiagonalizable weakly nonlinear
homogeneous system of hydrodynamic type. In fact, integrability
of the system (10.7) follows from Dubrovin's results [55] but
for the representation (10.7) it can be proved directly by
usual Hamiltonian and differential geometric methods of
hydrodynamic type systems.

\noindent{\bf Theorem 10.1 [58]}. The equation (10.3) is equivalent
to the integrable nondiagonalizable weakly nonlinear homogeneous
system of hydrodynamic type (10.7).

Analogously, for $N=3$ and for the second type special solutions (10.2)
of the WDVV system the associativity condition for algebra $A_t$
gives the following Dubrovin equation for the function
$f(x,y)$ [54]:

$$f_{xxx}f_{yyy} - f_{xxy}f_{xyy} =1 \eqno{(10.8)}$$

After introducing new variables $a,b,c,d$ (10.4)
from (10.8) we have the relation
$$d ={1+bc \over a}          \eqno{(10.9)}$$

Compatibility conditions (10.5) and the relation (10.9)
generate the following homogeneous system of hydrodynamic type
$${\pmatrix {a \cr
b \cr
c \cr}}_y = {\pmatrix {0 & 1 & 0 \cr
0 & 0 & 1 \cr
-{(1+bc) \over a^2} & {c \over a} & {b \over a} \cr}}
{\pmatrix {a \cr
b \cr
c \cr}}_x \eqno{(10.10)}$$

\noindent{\bf Theorem 10.2 [58]}. The equation (10.8) is
equivalent to the integrable nondiagonalizable weakly nonlinear
homogeneous system of hydrodynamic type (10.10).
\vskip .3in

This work was partially supported by the Russian Foundation
of Fundamental Researches (Grant No. 94-01-01478) and
the International Science Foundation (Grant No. RKR000).
\vskip .5in
\noindent{\bf References}

\vskip .2 in
\noindent
\item{[1]} B.A.Dubrovin and S.P.Novikov, Hydrodynamics of soliton
lattices, Sov. Sci. Rev. C, Math. Phys. 9 (1993), part 4, 1--136.
\item{[2]} B.A.Dubrovin and S.P.Novikov, Hamiltonian formalizm
of one-dimensional systems of hydrodynamic type and the
Bogolyubov-Whitham averaging method, Dokl. Akad. Nauk SSSR, 270,
No. 4 (1983) 781--785; (Soviet Math.Dokl., 27 (1983) 665--669).
\item{[3]} B.A.Dubrovin and S.P.Novikov, On Poisson brackets
of hydrodynamic type, Dokl. Akad. Nauk SSSR, 279, No. 2 (1984)
294--297; (Soviet Math. Dokl., 30 (1984) 651--654).
\item{[4]} B.A.Dubrovin and S.P.Novikov, Hydrodynamics of weakly
deformed soliton lattices. Differential geometry and Hamiltonian
theory, Usp. Mat. Nauk, 44, No. 6 (1989) 29--98; (Russian Math.
Surveys, 44, No. 6 (1989) 35--124).
\item{[5]} S.P.Tsarev, On Poisson brackets and one-dimensional
Hamiltonian systems of hydrodynamic type, Dokl. Akad. Nauk SSSR,
282, No. 3 (1985) 534--537; (Soviet Math. Dokl., 31 (1985) 488--491).
\item{[6]} S.P.Tsarev, Geometry of Hamiltonian systems of
hydrodynamic type. The generalized hodograph method, Izvestiya
Akad. Nauk SSSR, Ser. Mat., 54, No. 5 (1990) 1048--1068;
(Math. USSR Izvestiya, 37, No. 2 (1991)
397--419).
\item{[7]} O.I.Mokhov, Poisson brackets of Dubrovin-Novikov type
(DN-brackets), Funkts. Analiz i ego Prilozh., 22, No. 4 (1988) 92--93;
(Functional Anal. Appl., 22 (1988) 336--338).
\item{[8]} O.I.Mokhov, Symplectic forms on loop space and
Riemannian geometry, Funkts. Analiz i ego Prilozh. 24, No.3 (1990),
86--87;(Functional Anal. Appl., 24 (1990)).
\item{[9]} O.I.Mokhov, Homogeneous second order symplectic structures on
loop spaces and symplectic connections, Funkts. Analiz i ego Prilozh.,
25, No. 2 (1991) 65--67; (Functional Anal. Appl., 25 (1991)).
\item{[10]} O.I.Mokhov, Canonical Hamiltonian representation of the
Krichever-Novikov equation, Mat. Zametki, 50, No. 3 (1991) 87--96;
(Math. Notes  939--945).
\item{[11]} O.I.Mokhov, Two-dimensional nonlinear sigma models and
symplectic geometry on loop spaces of (pseudo)Riemannian manifolds,
Nonlinear Evolution Equations and Dynamical Systems. Proceedings
of the 8th Internat. Workshop (NEEDS'92), 6--17 July, 1992, Dubna,
Russia, Ed. V.G.Makhan'kov, World Sci. Publishing,
Singapore, 1993, 444--456
\item{[12]} I.Ya.Dorfman and O.I.Mokhov, Local symplectic operators
and structures related to them, J. Math. Physics, 32, No. 12 (1991)
3288--3296
\item{[13]} O.I.Mokhov, Geometry of commuting differential operators
of rank 3 and Hamiltonian flows, Ph.D. Thesis, Moscow State University,
Moscow, Russia, 1984.
\item{[14]} O.I.Mokhov, Symplectic structures on loop spaces of smooth
manifolds and Lagrangian systems of the field theory, Abstracts of
the Internat. Geom. Colloquium, May 10--14, 1993, Moscow, Russia, 38--39
\item{[15]} O.I.Mokhov and E.V.Ferapontov, On the nonlocal Hamiltonian
hydrodynamic type operators connected with constant curvature metrics,
Usp. Mat. Nauk, 45, No. 3 (1990) 191--192
\item{[16]} O.I.Mokhov, Hamiltonian systems of hydrodynamic type
and constant curvature metrics, Phys. Letters, 166A, No. 3--4 (1992)
215--216
\item{[17]} E.V.Ferapontov, Differential geometry of nonlocal
Hamiltonian operators of hydrodynamic type, Funkts. Analiz i ego
Prilozh., 25, No. 3 (1991) 37--49
\item{[18]} S.P.Novikov, Andrejewski Lectures, Berlin,
November-December 1993, Sfb 288 Preprint No. 117, 42 p.
\item{[19]} M.V.Pavlov, Multihamiltonian structures of
the Whitham equations, Dokl. Akad. Nauk, 338, No. 2 (1994) 165--167
\item{[20]} V.V.Sokolov, On Hamiltonian property of the Krichever-Novikov
equation, Dokl. Akad. Nauk SSSR, 272, No. 1 (1984) 48--50
\item{[21]} I.M.Krichever and S.P.Novikov, Holomorphic bundles and
nonlinear equations. Finite-zone solutions of rank 2, Dokl. Akad. Nauk
SSSR, 247, No. 1 (1979) 33--37
\item{[22]} I.M.Krichever and S.P.Novikov, Holomorphic bundles over
algebraic curves and nonlinear equations, Usp. Mat. Nauk, 35, No. 6
(1980) 47--68
\item{[23]} S.P.Novikov, Two-dimensional Schr\"odinger operators
in periodic fields, Itogi Nauki i Tekhn., Seriya  Sovremennye
Problemy Matematiki, 23 (1983) 3--32
\item{[24]} O.I.Mokhov, Commuting ordinary differential operators
of rank 3 corresponding to an elliptic curve, Usp. Mat. Nauk,
37, No. 4 (1982) 169--170
\item{[25]} O.I.Mokhov, Commuting differential operators of rank 3
and nonlinear equations, Izv. Akad. Nauk SSSR, Ser. Mat., 53, No. 6
(1989) 1291--1315
\item{[26]} S.I.Svinolupov and V.V.Sokolov, Evolution equations
with nontrivial conservation laws, Funkts. Analiz i ego Prilozh.,
16, No. 4 (1982) 86--87
\item{[27]} S.I.Svinolupov, V.V.Sokolov and R.I.Yamilov, B\"acklund
transformations for integrable evolution equations, Dokl. Akad. Nauk
SSSR, 271, No. 4 (1983) 802--805
\item{[28]} I.Ya.Dorfman, Krichever-Novikov equation and local
symplectic structures, Dokl. Akad. Nauk SSSR, 302, No. 4 (1988) 792--795
\item{[29]} I.Ya.Dorfman, Dirac structures of integrable evolution
equations, Phys. Letters, 125 A, No. 5 (1987) 240--246
\item{[30]} I.Ya.Dorfman, Dirac structures and integrability of
nonlinear evolution equations, Wiley, England, 1993
\item{[31]} G.Wilson, On the quasi-hamiltonian formalism of the KdV
equation, Phys.Letters, 132 A, No. 8--9 (1988) 445--450
\item{[32]} A.P.Fordy, in: Nonlinear Evolution Equations and
Dynamical Systems, Proc. of the 8th Internat. Workshop (NEEDS'92),
6--17 July, 1992, Dubna, Russia, Ed. V.G.Makhan'kov, World
Sci. Publishing, Singapore, 1993;
M.Antonowicz and A.P.Fordy, Reports in Math. Physics, 1993
\item{[33]} G.Reeb, Comptes Rendus, 229, No. 20 (1949) 969--971
\item{[34]} G.Reeb, Bull. Cl. Sciences, Acad. Royale Belg.,
5 S\'erie, 36, No. 4 (1950) 324--329
\item{[35]} A.Weinstein, J. Diff. Geom., 9, No. 4 (1974) 513--517
\item{[36]} A.Besse, Manifolds with closed geodesics, Mir, Moscow,
1981
\item{[37]} F.Magri, A simple model of the integrable Hamiltonian
equation, J. Math. Physics, 19, No. 5 (1978) 1156--1162
\item{[38]} O.I.Mokhov, Symplectic geometry on loop spaces of
smooth manifolds and nonlinear systems, Internat. Workshop
"Theory of Nonlinear Waves", September 1991, Kaliningrad University,
Kaliningrad, Russia
\item{[39]} O.I.Mokhov, Symplectic forms on loop spaces of
Riemannian manifolds, Internat. Conference "Differential equations
and related problems" in honour of 90-th anniversary of I.G.Petrovsky
(1901--1973), May 1991, Moscow State University, Moscow, Russia
\item{[40]} O.I.Mokhov, Two-dimensional $\sigma$-models in the
field theory: symplectic approach, Abstracts of the 9th Workshop
"Modern Group Analysis. Methods and Applications", June 1992,
Nizhniy Novgorod, Russia
\item{[41]} A.G.Meshkov, Hamiltonian and recursion operators for
two-dimensional scalar fields, Phys. Letters, 170 A, No. 6 (1992)
405--408
\item{[42]} C.Rogers, Reciprocal transformations and their
applications, in: Nonlinear Equations, Proc. 5th Workshop on Nonlinear
Evolution  Equations and Dynamical Systems (NEEDS'87), France, 1987,
109--123
\item{[43]} O.I.Mokhov and Y.Nutku, Bianchi transformation between
the real hyperbolic Monge-Amp\`ere equation and the Born-Infeld
equation, Letters in Math. Phys., 32, No. 2 (1994) 121--123
\item{[44]} G.V.Potemin, On Poisson brackets of
differential-geometric type, Dokl. Akad. Nauk SSSR, 286, No. 1 (1986)
39--42; (Soviet Math. Dokl., 33 (1986) 30--33).
\item{[45]} G.V.Potemin, Ph.D. Thesis, Moscow State University, Moscow,
Russia
\item{[46]} P.W.Doyle, Differential geometric Poisson bivectors
in one space variable, J. Math. Phys., 34, No. 4 (1993) 1314--1338
\item{[47]} S.P.Tsarev, Mat. Zametki, 46, No. 1, (1989) 105--111
\item{[48]} O.I.Mokhov, On Hamiltonian structure of evolution with
respect to the space variable $x$ for the Korteweg-de Vries equation,
Usp. Mat. Nauk, 45, No. 1 (1990) 181--182
\item{[49]} O.I.Mokhov, Joint Hamiltonian representation of the
Korteweg-de Vries equation and the three-wave equation, 1989
\item{[50]} Y.Nutku and \"O.Sar$\imath$o$\breve{g}$lu,
An integrable family of Monge-Amp\`ere equations and their
multi-Hamiltonian structure, Phys. Letters, 173 A, No. 3 (1993) 270--274
\item{[51]} B.L.Rozhdestvensky, N.N.Yanenko, Systems of quasilinear
equations and their applications to gas dynamics, Nauka, Moscow, 1978
\item{[52]} O.I.Mokhov and E.V.Ferapontov, Hamiltonian pairs associated
with
skew-symmetric Killing tensors on spaces of constant curvature,
Funkts. Analiz i ego Prilozh., 28, No. 2 (1994) 60--63
\item{[53]} O.I.Mokhov, Killing-Poisson bivectors on Riemannian
manifolds and integrable systems, Abstracts of Internat. Congress
of Mathematicians, 3--11 August 1994, Z\"urich, Switzerland, 50
\item{[54]} B.A.Dubrovin, Geometry of 2D topological field theories,
Preprint SISSA--89/94/FM, 1994, 204 p.
\item{[55]} B.A.Dubrovin, Integrable systems in topological
field theory, Nucl. Physics B, 379 (1992) 627--689
\item{[56]} E.Witten, On the structure of the topological
phase of two-dimensional gravity, Nucl. Physics B, 340 (1990)
281--332
\item{[57]} R.Dijkgraaf, E.Verlinde and H.Verlinde,
Nucl. Physics B, 352 (1991) 59;
Notes on topological string theory and 2D quantum gravity,
Preprint PUPT--1217, IASSNS-HEP--90/80, November, 1990
\item{[58]} O.I.Mokhov, Differential equations of associativity
in 2D topological field theories and geometry of nondiagonalizable
systems of hydrodynamic type, Abstracts of Internat. Conference on
Integrable Systems "Nonlinearity and Integrability: from Mathematics
to Physics", February 21--24, 1995, Montpellier, France
\item{[59]} Yu.N.Sidorenko, Zap. Nauchn. Sem. LOMI, 161, No. 7 (1987)
76--87
\item{[60]} E.Barouch, A.S.Fokas and V.G.Papageorgiou, Journal of
Math. Physics, 29, No. 12 (1988) 2628--2633
\item{[61]} I.M.Gelfand and I.Ya.Dorfman, Hamiltonian operators
and algebraic structures related to them, Funkts. Analiz i ego Prilozh.,
13, No. 4 (1979) 13--30;
I.M.Gelfand and I.Ya.Dorfman, Schouten bracket and Hamiltonian
operators, Funkts. Analiz i ego Prilozh., 14, No. 3 (1980) 71--74
\item{[62]} A.S.Fokas and B.Fuchssteiner, On the structure of
symplectic operators and hereditary symmetries, Lettere al Nuovo
Cimento, 28, No. 8 (1980) 299--303;
B.Fuchssteiner and A.S.Fokas, Symplectic structures, their B\"acklund
transformations and hereditary symmetries, Physica D, 4 (1981) 47--66
\item{[63]} I.Ya.Dorfman and F.W.Nijhoff, On a (2+1)-dimensional
version of the Krichever-Novikov equation, Phys. Letters A, 157 (1991)
107--112
\bye